\documentclass[twocolumn, prl, superscriptaddress, amsmath, amssymb, aps]{revtex4-2}
\usepackage{multirow}
\usepackage{siunitx} 
\usepackage{amsmath}
\usepackage{amssymb}
\usepackage{braket}
\usepackage{mathtools}
\usepackage{graphicx}
\usepackage{color}
\usepackage{xcolor}
\usepackage{ulem}
\usepackage{xspace}
\usepackage{lipsum}
\usepackage{upquote}
\usepackage{lipsum}
\usepackage{subcaption} 
\usepackage{ragged2e} 
\DeclareCaptionJustification{justified}{\justifying} 
\usepackage[margin=1 in]{geometry}
\usepackage[utf8]{inputenc}
\usepackage{csquotes}
\usepackage{sansmathaccent}
\usepackage{tikz}
\usepackage{minitoc}
\usepackage[bookmarks=true,colorlinks=true,urlcolor=blue,linkcolor=blue,citecolor=blue,breaklinks]{hyperref}
\usepackage{hyperref}
\usepackage{tocloft}
\usepackage{float}
\usepackage{color}
\hypersetup{colorlinks= true, citecolor= blue}

\newcommand{\bfr}{\mathbf{r}}
\newcommand{\bfk}{\mathbf{k}}
\newcommand{\bfG}{\mathbf{G}}
\newcommand{\bfq}{\mathbf{q}}

\newcommand{\moire}[0]{moir\'{e}\xspace}
\newcommand{\vF}[0]{v_\mathrm{F}}
\newcommand{\kF}[0]{k_\mathrm{F}}
\newcommand{\EF}[0]{E_\mathrm{F}}
\newcommand{\figref}[1]{Fig.~\ref{#1}}
\newcommand{\eqnref}[1]{Eq.~(\ref{#1})}
\newcommand{\BZ}{Brillouin zone}
\DeclareUnicodeCharacter{0650}{\unskip}
\begin{document}

\title{A self-consistent Hartree theory for lattice-relaxed magic-angle twisted bilayer graphene}

\author{Mohammed M. Al Ezzi}
\affiliation{Department of Materials Science and Engineering, 
National University of Singapore, 9 Engineering Drive 1, 
Singapore 117575}
\affiliation{Department of Physics, Faculty of Science, National University of Singapore, 2 Science Drive 3, Singapore 117542}
\affiliation{Centre for Advanced 2D Materials, National University of Singapore, 6 Science Drive 2, Singapore 117546}
\author{Liangtao Peng}
\affiliation{Department of Physics, Faculty of Science, National University of Singapore, 2 Science Drive 3, Singapore 117542}
\affiliation{Centre for Advanced 2D Materials, National University of Singapore, 6 Science Drive 2, Singapore 117546}
\author{Zhengyu Liu}
\affiliation{Centre for Advanced 2D Materials, National University of Singapore, 6 Science Drive 2, Singapore 117546}
\author{Jonah Huang Zi Chao}
\affiliation{Department of Physics, Faculty of Science, National University of Singapore, 2 Science Drive 3, Singapore 117542}
\author{Gayani N. Pallewela}
\affiliation{Centre for Advanced 2D Materials, National University of Singapore, 6 Science Drive 2, Singapore 117546}
\author{Darryl Foo}
\affiliation{Centre for Advanced 2D Materials, National University of Singapore, 6 Science Drive 2, Singapore 117546}
\author{Shaffique Adam}
\affiliation{Department of Materials Science and Engineering, 
National University of Singapore, 9 Engineering Drive 1, 
Singapore 117575}
\affiliation{Department of Physics, Faculty of Science, National University of Singapore, 2 Science Drive 3, Singapore 117542}
\affiliation{Centre for Advanced 2D Materials, National University of Singapore, 6 Science Drive 2, Singapore 117546}
\affiliation{Department of Physics and Astronomy, University of Pennsylvania, Philadelphia, Pennsylvania 19104, USA}
    
\date{\today} 

\begin{abstract}
\normalsize
For twisted bilayer graphene close to magic angle, we show that the effects of lattice relaxation and the Hartree interaction both become simultaneously important.  Including both effects in a continuum theory reveals a Lifshitz transition to a Fermi surface topology that supports both a ``heavy fermion" pocket and an ultraflat band ($\approx 8~{\rm meV}$) that is pinned to the Fermi energy for a large range of fillings.  We provide analytical and numerical results to understand the narrow ``magic angle range" that supports this pinned ultraflat band and make predictions for its experimental observation.  We believe that the bands presented here are accurate  at high temperature and provide a good starting point to understand the myriad of complex behaviour observed in this system.
\end{abstract}
\maketitle

\doparttoc 
\faketableofcontents 
\part{} 

\emph{\it Introduction} ---  It has now been six years since the seminal experimental observations of superconductivity~\cite{cao2018unconventional} and correlated insulators~\cite{cao2018correlated} in magic-angle twisted bilayer graphene.  The presence of flat bands at magic angle was anticipated within the continuum theory for moiré quantum materials~\cite{dos2007graphene, bistritzer2011moire,dos2012continuum}.  Nonetheless, there is a wide consensus among the theoretical community that the non-interacting bands of the original continuum model are a poor starting point to describe experiments close to magic angle. This might be surprising because of the remarkable success that the continuum model has seen at larger twist angles \cite{wong2015local}. 

\begin{figure}[t!]
\includegraphics[width=0.9\linewidth, height=0.7\linewidth]{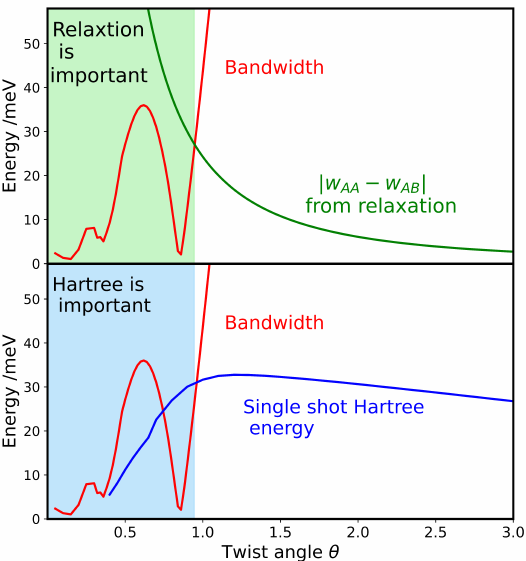}
\caption{For twisted bilayer graphene, relaxation and Hartree interaction corrections become simultaneously important close to magic angle.  We compare the relaxation and Hartree interaction energy scales to the non-interacting bandwidth as a function of twist angle. Lattice relaxation and Hartree interactions both modify the band structure, with their effects becoming more relevant as bandwidth decreases.  To obtain an accurate model near magic angle, both effects must be taken into account.}
\label{Fig:Schematic}
\vspace{-0.2in}
\end{figure}

So what goes wrong with the continuum model?  As shown schematically in Fig.~\ref{Fig:Schematic}, two effects become simultaneously important as one approaches the magic angle.  First, at low twist angle, small displacements in the atomic positions give rise to large gains in electronic energy.  This is known as ``lattice relaxation” and has been appreciated using very different techniques that largely agree \cite{nam2017lattice,carr2019exact,guinea2019continuum,cantele_structural_2020,leconte2022relaxation, bennett2023twisted, ceferino2023pseudomagnetic, ezzi2023analytical,pan2024structural}. Second, the \moire potential hybridizes the wavefunctions from each layer creating excess charge density at the ``AA regions" where each carbon atom in one layer is directly above one in the other layer.  This Hartree correction strongly modifies the continuum electronic bands destroying the flat bands \cite{guinea_electrostatic_2018, rademaker_charge_2019, cea2019electronic,xie2020nature, goodwin_hartree_2020}. To our knowledge, the effect of the Hartree interaction and accurate lattice relaxation have not both been systematically included in the theory.  

\begin{figure*}[t!]
\includegraphics[width= \linewidth]{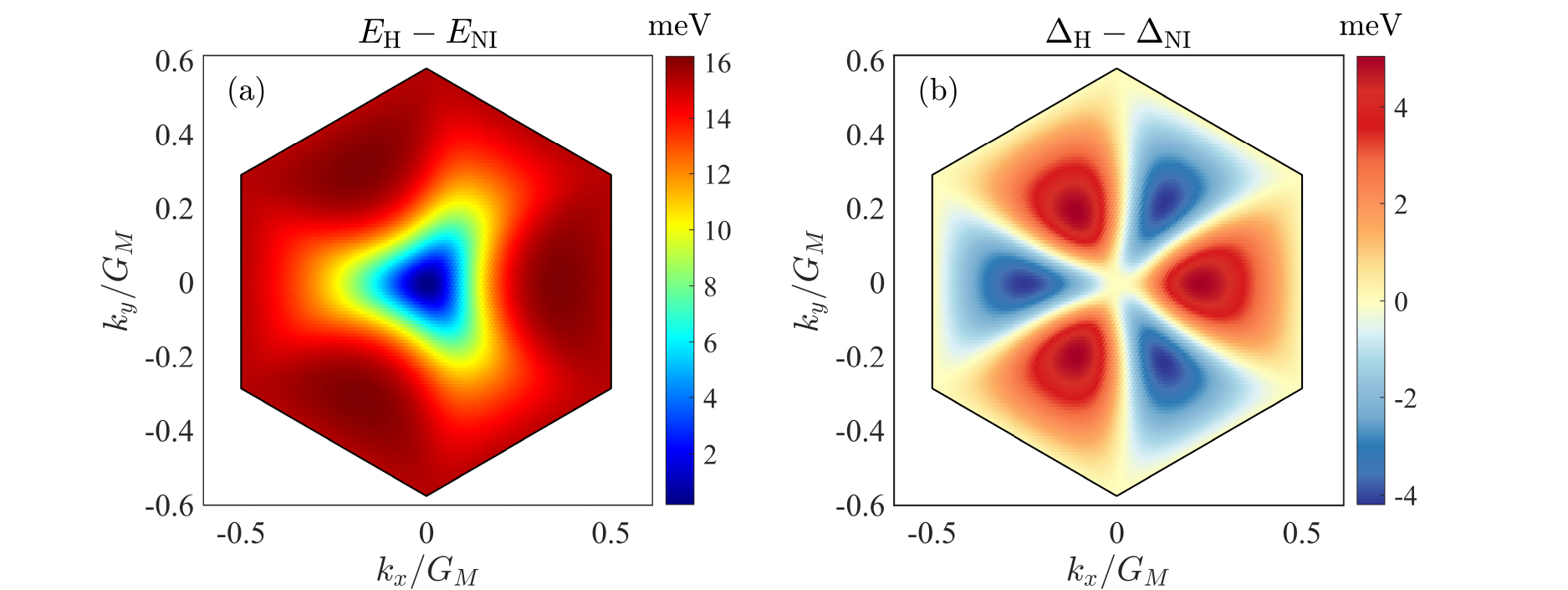}
\caption{Two visualizations of the effect of the Hartree interaction correction. (a)  Energy shift of the conduction band due to interactions, measured as the difference 
$E_\mathrm{H} - E_\mathrm{NI}$ between the interacting and non-interacting energies at $\theta=1.05^\circ$ and $\nu=2$.  The $\Gamma$ point (centre) remains fixed with the rest of the band showing a sizable upward shift in energy.  From this perspective, the Hartree interaction significantly distorts the bands.  (b) Degree of band shape distortion $\Delta_\mathrm{H} - \Delta_\mathrm{\rm NI}$ measured as the difference in the direct gap $\Delta$ between valence and conduction bands both with and without the Hartree interaction.  The pale yellow regions on the K--M--K$'$ border and at the $\Gamma$ point show where the band shapes are locally undistorted, undergoing only a rigid shift.  As explained in the text, this rigid shift explains why the Hartree interaction would not be visible in common experiments like transport at the Dirac point or scanning probe measurements.}
\label{Fig:ShiftOfBands}
\end{figure*}

We accomplish this in this work and find a new and remarkable Lifshitz transition as a function of filling.  The fully relaxed theory has a new Fermi surface topology that supports a ``heavy fermion" electron pocket at the $\Gamma$-point that is accompanied by an ultraflat band with bandwidth $w \approx\!8$ meV that is pinned above the Fermi energy for a large fraction of the band filling $\Delta \nu \approx\!40$ percent.  This causes a kink in the energy as a function of filling that could be probed directly in compressibility measurements, or indirectly in  transport experiments. The Lifshitz transition is only present when both self-consistency and lattice relaxation are included in the theory and it is only stable for a narrow range of twist angles $\Delta \theta \approx\! 0.2$ degrees that roughly coincides with the original magic angle of the non-interacting rigid continuum model.  We therefore understand the emergence of heavy fermions in magic-angle twisted bilayer graphene~\cite{song2022magic, rozen_entropic_2021} as a consequence of combining mean-field electrostatics and lattice relaxation rather than any exotic and mysterious correlation effect.

 The rigid lattice approximation is valid at large twist angles.  Here we have $W_{AA} = W_{AB}$, where these are the interlayer hopping elements between sublattice A in one layer and either sublattice $A$ or $B$ in the second layer, respectively.  Lattice relaxation can be captured by the quantity $|W_{AA} - W_{AB}|$ that increases at smaller angles and becomes comparable to the bandwidth for $\theta \lesssim 1$ degree.  Lattice relaxation makes several qualitative changes to the band structure including isolating the flat bands from higher energy bands \cite{nam2017lattice,carr2019exact}, shifting the numerical value of the magic angle from the rigid lattice prediction of $\theta_M \approx 1.05$ degrees \cite{carr2019derivation,leconte2022relaxation}, the non-vanishing of the Fermi velocity creating a ``magic-range”  \cite{bennett2023twisted,ezzi2023analytical}, and generating large pseudomagnetic fields \cite{ezzi2023analytical,ceferino2023pseudomagnetic}.  We note that while relaxation shifts the ``magic range" to higher values, including the effects of the resultant pseudomagnetic fields roughly shifts it back close to the original value.

Similarly, the Hartree interaction that modifies the electronic bands to  redistribute the excess charge caused by the \moire potential vanishes at large twist angle. The single-shot Hartree energy (defined below) coincidentally becomes comparable to bandwidth also around $\theta \approx 1^\circ$.  The Hartree correction destroys the flat bands giving bandwidths set instead by the Hartree energy.  This is most evident in the chiral model where $W_{AA} = 0$.  Without Hartree, the chiral model exhibits a doubly degenerate perfectly flat non-dispersing band. However, including the Hartree interaction yields a parabolic dispersing band centered at the $\Gamma$ point and a flattish band around the K-M regions of the \BZ.  These two regions are separated in energy by the Hartree energy.  This picture is qualitatively the same for the full model except that the Dirac fermions at the K-point remain dispersive with a twist-angle dependent renormalized Fermi velocity $v^*$ -- and even when $v^*$ vanishes at magic angle, the quadratic terms do not, and the band remains dispersive~\cite{hejazi_multiple_2019}.  

Unlike the relaxation problem where different numerical and analytical approaches give largely consistent conclusions for the lattice reconstruction, the literature on the Hartree problem is expansive and plagued with inconsistencies \cite{guinea_electrostatic_2018, rademaker_charge_2019, cea2019electronic,xie2020nature, goodwin_hartree_2020, zhang_correlated_2020, bultinck_ground_2020, lewandowski_does_2021, liu_theories_2021, bernevig_twisted_2021, song2022magic, wagner_global_2022}.  One byproduct of our work is that we carefully benchmark different approaches and check numerical convergence.  This allows us to make several theoretical statements including: (i) To calculate the Hartree interaction one should use the physically relevant dielectric constant of the substrate $\epsilon \approx 4$.  It is common in the literature to use unreasonably large values of $\epsilon$ to artificially suppress the role of the Coulomb interaction.  We reveal that the theoretical argumentation for this dielectric constant enhancement breaks down once more than 6 continuum bands are considered.  We show that once relaxation effects are included, there is no need to artificially suppress the Hartree interaction to get reasonable results; (ii) the single-shot Hartree approach and self-consistent Hartree only agree for $\epsilon \gtrsim 15$.  Indeed, we can prove that self-consistency is always important when the Hartree interaction energy scale is comparable to bandwidth; and (iii) we show analytically for the chiral model that the sum of the electron potential from the bottom of the bands to charge neutrality vanishes (see also Ref. \cite{becker2023exact}), and that this result also holds numerically for the realistic model.  This implies that single-shot Hartree results will be identical regardless of the so-called ``subtraction scheme" employed i.e. whether one sums the potential from the bottom of the band, or from charge neutrality.  However, the necessity of self-consistency also imposes that we must sum the potential over all the filled bands.  These technical points are left to the Supplemental Material. \\

\noindent \emph{{\it Model}}--- To keep the model parameters to a minimum while preserving the high accuracy of the theory, we choose to neglect out-of-plane relaxation. We have checked numerically that this approximation does not change any of our conclusions and makes only small quantitative changes to our results.  However, this then allows us to write down the complete theory with only 6 highly constrained parameters.  These are all set to carefully calibrated values well within the range of values accepted in the literature. Our model parameters are: monolayer graphene Fermi velocity $v_{\rm F}= 1.05 \times 10^{6}~{\rm m/s}$ and lattice constant $a_0 = 0.246~{\rm nm}$, substrate dielectric constant $\epsilon =4$, interlayer \moire potential $W^{*} = 100~{\rm meV}$, $\Delta W(\theta) = A/\sin^2(\theta/2)$, with $A = 1.85 \times 10^{-3}~{\rm meV}$, where we define $W_{AB(A)}(\theta)\equiv W^* \pm  \Delta W(\theta)$; and pseudomagnetic field energy scale $\gamma(\theta) = B /\sin(\theta/2)$, where $B =9.5 \times 10^{-2}~{\rm meV}$.  We have a detailed discussion of the choice of model parameters in the Supplemental Material, but mention here that following Ref.~\cite{ezzi2023analytical} the values of $A$ and $B$ are derived from the relaxed atomic positions obtained from LAMMPS molecular dynamics simulations.  We have made our codes publicly available so that others can use our model~\footnote{\url{https://github.com/LiangtaoPeng/TBGHartree}}.  

Our noninteracting twisted bilayer graphene Hamiltonian takes the familiar form
\begin{align}
    H_{0}(\bfr)=
    \begin{pmatrix}
    D_1 & T(\bfr)\\
    T^\dagger(\bfr) & D_2
    \end{pmatrix}_{L},
    \label{Eq:noninteracting}
\end{align}
where $D_l$ is the monolayer Dirac cone for layer $l$, $T(\bfr)$ is the \moire potential and the subscript $L$ denotes that we are in layer space. The Dirac cones are shifted by a pseudomagnetic field generated by in-plane relaxation of atomic positions, 
\begin{align}
    D_l=\boldsymbol{\sigma}_{\theta/2\pm}\cdot\left(-i\hbar\vF\nabla\pm\mathbf{b}_\mathrm{ps}(\bfr,\theta)\right),\\
    \mathbf{b}_\mathrm{ps}(\bfr,\theta)=\gamma(\theta)(\mathrm{Re}\{\tilde{\partial} \tilde{D}(\bfr)\},\mathrm{Im}\{\tilde{\partial} \tilde{D}(\bfr)\}),
\end{align}
where the sublattice Pauli matrices are rotated by $\theta/2$, $\sigma^{x,y}_{\theta/2}=e^{-(i\theta/4)\sigma^z}\sigma^{x,y}e^{(i\theta/4)\sigma^z}$ reflecting the twists of both layers, 
$\tilde{D}\equiv D_x+iD_y$ is a phasor representation of the atomic relaxation displacement field $\mathbf{D}=(D_x,D_y)$ and $\tilde{\partial}\equiv\partial_x+i\partial_y$ is the chiral space derivative. The interlayer \moire potential is

\begin{align}
    T(\bfr)=\sum_{j=0}^{2}e^{i\mathbf{g}_j\cdot\bfr}
    \begin{pmatrix}
        W_{AA} & W_{AB}e^{-2\pi i j/3}\\
        W_{AB}e^{2\pi i j/3} & W_{AA}
    \end{pmatrix}_{S},
\end{align}
where $\mathbf{g}_0=\frac{8\pi}{3a_0}\sin(\frac{\theta}{2})(0,1)$ and $\mathbf{g}_j=R^j_{2\pi/3}\mathbf{g}_0$ with $R_\phi$ the rotation operator for an angle $\phi$ are the three vectors connecting a \moire K point to the three nearest K$'$ points, and the subscript $S$ denotes sublattice space. Spin and valley labels contributing to the 4-fold degeneracy are omitted for brevity. 

We consider only the Hartree interaction and neglect the spin and valley exchange splitting of the \moire bands.  The exchange energy at the monolayer graphene level is already incorporated into the renormalized Fermi velocity as discussed in Ref.~\cite{tang2018role}.  Since the \moire-induced exchange splitting results in small energy differences $\sim3~{\rm meV}$ \cite{zhu2024gw}, the present work should be thought of as accurate for experiments done at temperatures above $35~{\rm K}$.  The Hartree interaction Hamiltonian is given by
\begin{align}
    H_\mathrm{Har}(\bfr)=4\int\mathrm{d}^2\bfr'U(\bfr-\bfr')(n(\bfr')-n_0),
\end{align}

\noindent where $U(\bfr)$ is the gate-screened Coulomb interaction, $n$ is the self-consistently determined electron density, and the factor of 4 accounts for spin and valley degeneracy.
$n_0$ is the constant average electron density that gets cancelled by the positively charged nuclei.
Since $H_\mathrm{Har}$ is diagonal in layer and sublattice space, Fourier transforming gives
\begin{align}
    H_{\mathrm{Har}}(\bfk)=4\sum_{\bfG\neq0}U_\bfG\rho_\bfG \psi^\dagger_\bfk \psi^{}_{\bfk-\bfG}.
\end{align}
The $\bfG$ are \moire reciprocal lattice vectors with $\bfG=0$ omitted due to the subtraction of $n_0$.  The operators $\psi^\dagger_\bfk$ and $\psi_\bfk$ create and annihilate electrons of momentum $\bfk$, respectively, where the layer and sublattice labels are suppressed for brevity.   From the symmetry of the \moire lattice, we have $U_\bfG=U_G$ and $\rho_\bfG=\rho_G$, where $G=|\bfG|$.  We see that there are two contributions to the Hartree interaction:  First, the Fourier transform of the Coulomb potential which for a double-gate structure is

\begin{align}
U_G=\frac{2\pi}{A_m}\frac{e^2 \tanh (Gd)}{\epsilon G} \xrightarrow{d\rightarrow\infty} \frac{e^2 \theta}{\epsilon a_0} \approx 25~{\rm meV},
\end{align}
where $A_m$ is the area of a \moire unit cell and $d$ is the distance from each gate to the bilayer.  We show in the Supplemental Material that the value of $d$ is irrelevant above 2 nm; and since most experimental values are on the order of 10 nm, we can safely take $d\rightarrow \infty$ in everything that follows.  In the numerical estimate, we use a typical value of $\theta=1^\circ$.  The second contribution is the density harmonics $\rho_G=\sum_\bfk\langle \psi^\dagger_\bfk \psi^{}_{\bfk-\bfG}\rangle =  \sum_{n,k}  \Lambda^{nn}_{\bfk,\bfG}$ that we determine self-consistently. The form factor  
\begin{align}
    \Lambda^{nn'}_{\bfk,\bfG}=\sum_{\bfG'}c^*_{n,\bfk+\bfG'}c^{}_{n',\bfk+\bfG'+\bfG}.
\end{align}
\noindent 
is defined in terms of the Bloch coefficients $c^{}_{n,\bfk+\bfG}$ of $\psi(\bfr)$. To simplify the notation we introduce $\Lambda_\mathrm{k}=\Lambda_{\mathrm{k},G_0}$ where we drop the band indices superscripts of $\pm1$ that can be inferred from the context.  We reiterate that the self-consistency step is important whenever the Hartree correction itself is significant.  For the single-shot Hartree approximation, the density harmonics are computed considering only the non-interacting Hamiltonian in Eq.~\ref{Eq:noninteracting}. These definitions of $U_G$, $\rho_G$ and $\Lambda^{nn'}_{\bfk,\bfG}$ fully specify our Hartree interaction theory.  \\

\begin{figure*}[th!]
\includegraphics[width= 0.95\linewidth]{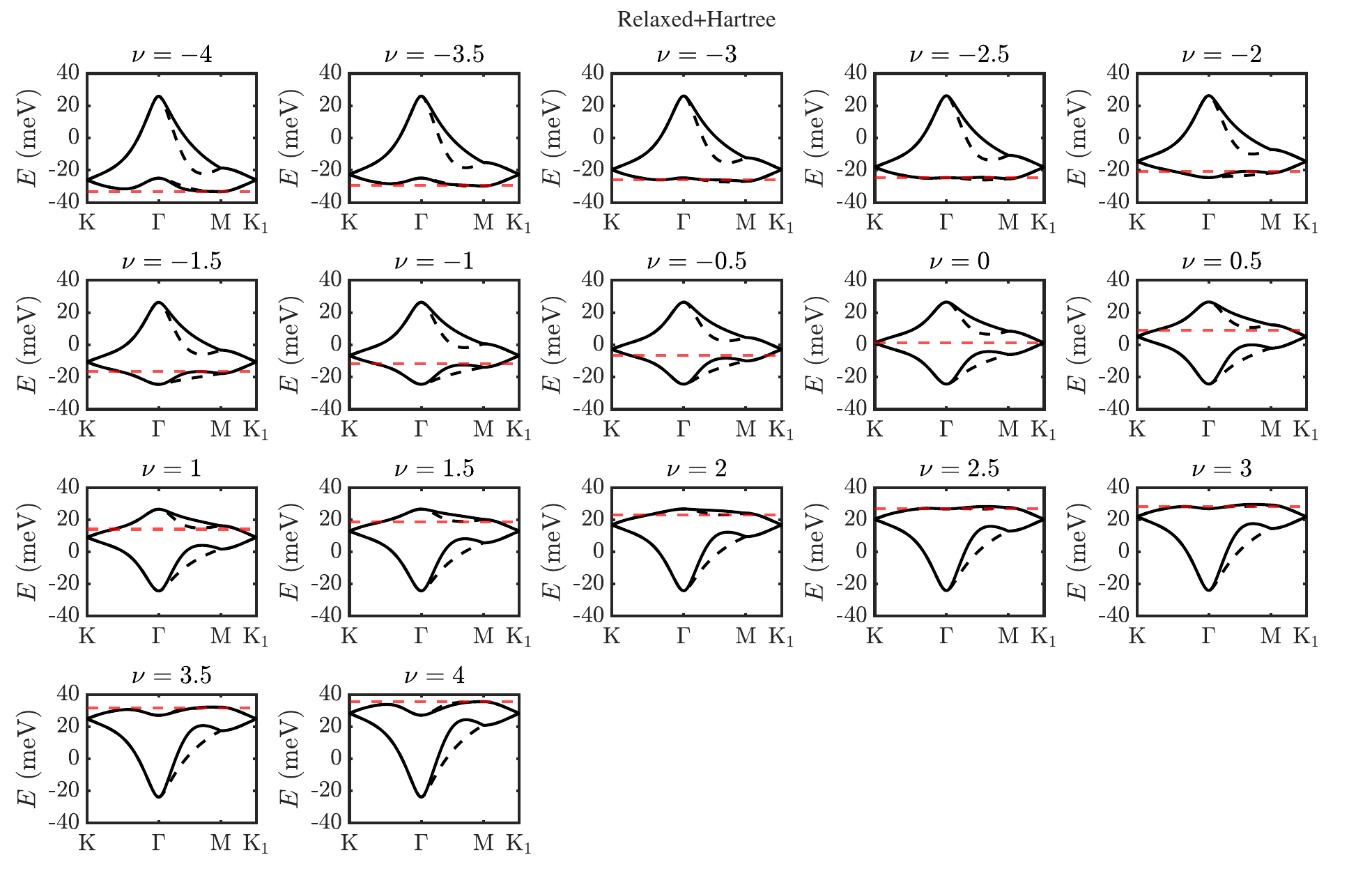}
\caption{Reconstruction of lattice relaxed bands with self-consistent Hartree interactions.  The low energy band starts filling at $\nu = -4$ from the M point.  The $K$-point is half-filled at charge neutrality $\nu=0$ where the Hartree correction vanishes and the bandstructure is identical to the non-interacting theory.  Notice that increasing $\nu>0$ the curvature of the conduction band at the $\Gamma$ point changes from negative at $\nu=0$ to positive at $\nu=4$ necessitating a diverging effective mass at $\nu_{\rm FB}$.  Our numerical and analytic results show that $\nu_{\rm FB} \approx \nu^*$ defined as the energy where $E_M(\nu^*) = E_\Gamma(\nu^*)$ implying an ultraflat band.  In the relaxed model the $\Gamma$ point filling $\nu_\Gamma \approx 2.5 > \nu^*$.  This implies a heavy fermion pocket at the $\Gamma$ point that stabilizes the ultraflat band.  The ultraflat band is pinned to the Fermi energy for $\Delta \nu \approx 40$ percent of the band filling.}
\label{Fig:Bands_with_EF_relax_Hartree}
\end{figure*}

\noindent \emph{\it Results}---  The first issue we address is why although the Hartree correction is as large as the bandwidth (see Fig.~\ref{Fig:Schematic}) its effects have not been readily observed in experiments. Figure \ref{Fig:ShiftOfBands} illustrates an important point.  The left panel shows the difference between the Hartree corrected bands and the non-interacting bands in the moire Brillouin zone.  It shows that the region around the $\Gamma$ point is least affected by Hartree while the region around the K and M points are strongly affected by Hartree.  As discussed in the Supplemental Material, this can be understood from the wavefunctions at $K$ and $\Gamma$.  But this is only part of the story.  In the right panel of Fig.~\ref{Fig:ShiftOfBands} we look at the energy difference between the first conduction and first valence bands (denoted as $\Delta$).  We compare $\Delta_H$ for the Hartree corrected bands with $\Delta_{\rm NI}$ for the non-interacting bands.  Here a very different picture emerges.  For all the high symmetry points $\Gamma$, $K$, and $M$, the Hartree correction just rigidly shifts the non-interacting bands without changing the underlying band structure. In the Supplemental Material, we show that properties like the Fermi velocity, bandwidth, and separation between the Van Hove singularities (that are close to the $M$-point) are largely unaffected by the Hartree interaction, and that this is true both with and without relaxation.  This explains why typical experimental probes like transport at the Dirac point \cite{polshyn2019large} are well described by non-interacting band structures \cite{sharma2021carrier}.  Similarly, scanning probe measurements \cite{choi2021interaction,wong2015local,li2010observation,kerelsky2019maximized,xie2019spectroscopic} that measure the distance between the Van Hove singularities would not detect the sizable Hartree interaction correction.

However, the Hartree does have important and observable consequences.  Unlike the non-interacting model where the bands start filling from the $\Gamma$-point, the Hartree corrected bands start filling from the $M$-point.   We denote the filling of the $\Gamma$-point as $\nu_\Gamma$.  At charge neutrality, the Hartree correction vanishes and the $K$-points are filled. The most obvious effect of the Hartree is the linear shift of the bands with filling.  We find that this linear regime extends for all fillings $|\nu| < |\nu_\Gamma|$.  We can expand the Hartree correction at momentum $k$ about $\nu=0$ to find

\begin{align}
E^{H}_{k}(\nu) = 6 \eta \nu U_G  \Lambda_\mathrm{K} \Lambda_k.
\label{Eq:perturbative}
\end{align}

\noindent The band reconstruction at point $k$ can understood as the product of five terms: the form factor at this momentum $\Lambda_k$, the excess charge given by the filling factor $\nu$, the interaction strength $U_G$, the form factor at charge neutrality $\Lambda_\mathrm{K}$, and the self-consistency factor $\eta$ that captures higher order processes.  The factor of 6 accounts for the symmetry-equivalent lattice vectors $\bfG$. For the single-shot Hartree calculation $\eta = 1$, while $\eta < 1$ for the self-consistent Hartree interaction.  The Supplemental Material contains the derivation of Eq.~\ref{Eq:perturbative} and the full expression for $\eta$.  For $\theta = 1.05^\circ$, Eq.~\ref{Eq:perturbative} gives $E^H(K) = E^H(M) = \left(25~{\rm meV} \right) \nu$ for $\eta = 1$, in exact agreement with our full numerics for single-shot Hartree in both the rigid and relaxed models.  However, we find that including self-consistency gives $\eta \approx 0.3$.

 The band evolution with filling $\nu$ is shown in Fig.~\ref{Fig:Bands_with_EF_relax_Hartree}.  We show a similar evolution for the rigid case in the Supplemental Material.  Since in the small twist angle regime, our bands are approximately symmetric for positive and negative filling, we discuss the the filling of the bands starting from charge neutrality and then increasing $\nu \geq 0$.  At $\nu=0$, we have the non-interacting conduction band that starts by filling the Dirac fermions at the K-point, with the global energy maximum at the $\Gamma$ point.  However, since
 as discussed in Fig.~\ref{Fig:ShiftOfBands}, $E^H_\Gamma(\nu) \sim \Lambda_\Gamma \approx 0$, the surrounding regions shift upwards with increased filling.  The effective mass $m^*$ at the $\Gamma$ point changes from negative to positive, diverging when the band is perfectly flat.  We define $\nu_{\rm FB}$ as the filling where $m^*=\infty$, and $\nu^*$ as the filling when $E_M(\nu) = E_\Gamma(\nu)$.  From our numerics and within our analytical model we can show that with self-consistency $\nu^* \approx \nu_{\rm FB}$.  This simultaneously satisfies two definitions of flat bands: a local definition looking at the diverging effective mass at the $\Gamma$-point, and a more global definition equating the energies at the $M$ and $\Gamma$ points.  The consequence is that the emergent ultraflat band induced by the self-consistent Hartree interaction occupies a large fraction of the \moire Brillouin zone.

 \begin{figure*}[ht!]
\includegraphics[width=0.9\linewidth]{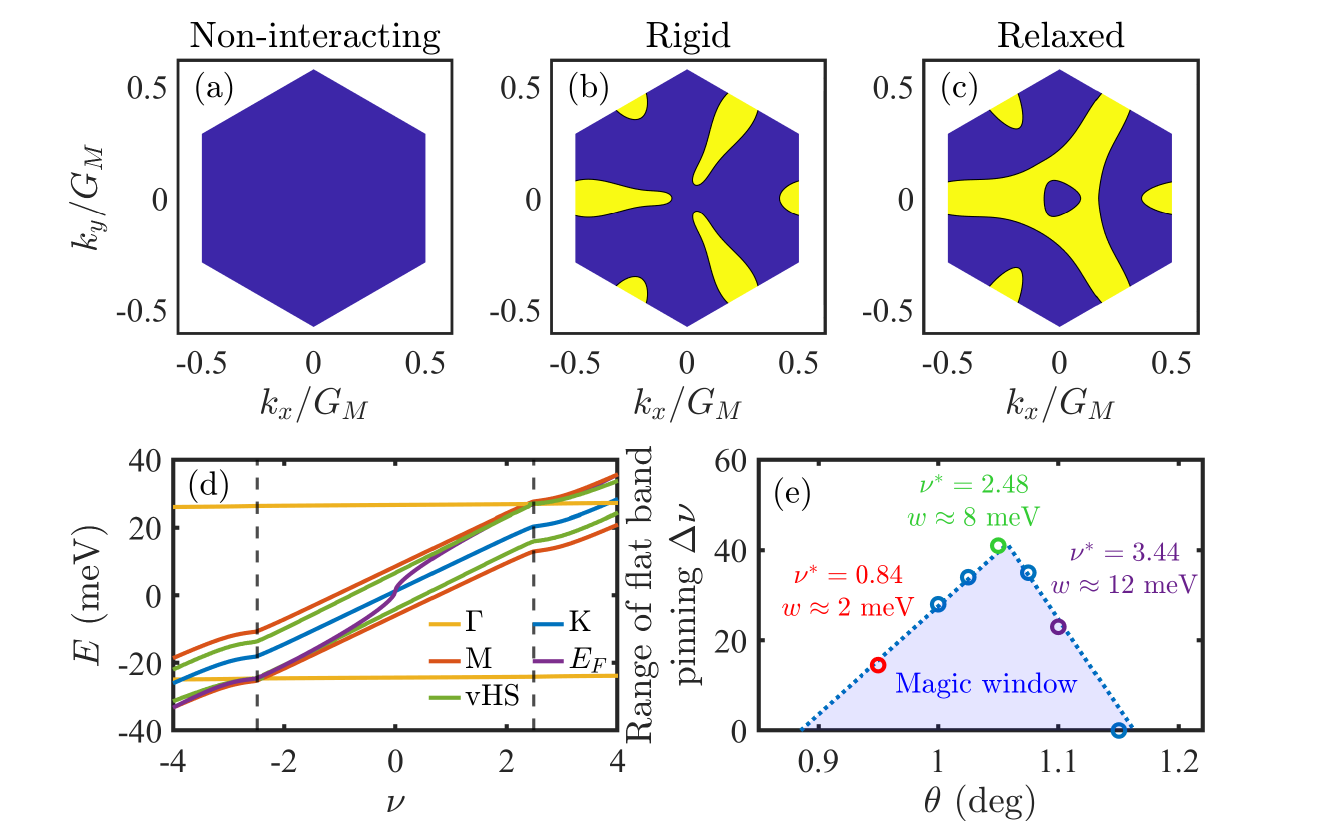}
\caption{Emergence of a heavy fermion and an ultraflat band.  Upper panels show the Fermi surface when the $\Gamma$-point is filled for $\theta=1.05^\circ$. (a) In the non-interacting model, the $\Gamma$-point is the last to fill. (b) For the rigid model with self-consistent Hartree, the $\Gamma$-point fills as part of a large connected Fermi surface.  (c) For the relaxed model with self-consistent Hartree, the $\Gamma$ point fills as an isolated fermion pocket that contributes very weakly to further renormalization of the bands.  (d) The energies at the high symmetry points together with $E_F$ and $E_{\rm vHS}$ as a function of filling.  The dashed line is $\nu_\Gamma$ defined as when $E_F = E_\Gamma$.  In the relaxed model we have the double coincidence $\nu_\Gamma \approx E_M \approx E_{\rm vHS}$ causing a kink in all energies except $E_\Gamma$.  (e) The ultraflat band only occurs in a narrow ``magic window" close to the magic angle.  At $\theta = 1.05$ the ultraflat band onset occurs at $\nu^* = 2.48$, has bandwidth of $8$ meV and is pinned for $\Delta \nu = 40$ percent of the band filling.  $\Delta \nu$ drops linearly on either side of this maximum.}
\label{Fig:FlatBand}
\vspace{-0.15in}
\end{figure*}

It is close to $\nu_\Gamma$ where the predictions of the rigid and relaxed models sharply diverge. We find that $\nu_\Gamma = \nu^* \pm \mathcal{O}(v^*/v_F)$, where the correction is negative for the rigid model and positive for the relaxed model.  This has significant implications on the Fermi surface topology.  For the rigid model, $\nu_\Gamma < \nu^* \approx \nu_{\rm FB}$, which implies that the $\Gamma$ point gets filled while it still has a negative curvature.  When the perfectly flat band emerges, it is already completely filled and there is no change in the Fermi surface topology.  There is negligible change in the linear-in-$\nu$ Hartree slopes at the $K$ and $M$ points obtained from Eq.~\ref{Eq:perturbative}, and the bands evolve continuously without anything interesting happening at $\nu_{\Gamma}$.  

\newpage 

For the relaxed theory, $\nu^* \approx \nu_{\rm FB} < \nu_{\Gamma}$.  This implies that when the perfectly flat band emerges, it is unoccupied.  There is then a Lifshitz transition at $\nu_\Gamma$ as the $\Gamma$-point starts to fill as a local minimum with positive curvature.  It forms a distinct heavy fermion electron pocket at the $\Gamma$ point.  The upper panel of Fig.~\ref{Fig:FlatBand} shows the Fermi surface topology close to $\nu_\Gamma$ contrasting the non-interacting theory, and self-consistent Hartree calculation for the rigid and relaxed models. Filling electrons into the heavy fermion pocket at the $\Gamma$-point results in little additional Hartree modification to the bands since $\Lambda_\Gamma \approx 0$.  This has two important physical consequences both shown in Fig.~\ref{Fig:FlatBand}.  First, there is a kink in total energy $E(\nu)$ going from linear slope to vanishing slope at $\nu_\Gamma$.  Second, the vanishing slope for $\nu > \nu_\Gamma$ persists over a wide range of band filling $\nu$.  While the pinning of the Van Hove singularity to the Fermi energy is not surprising and has been reported previously~\cite{cea2019electronic}, its coincidence with $\nu_\Gamma$ occurs only when both relaxation and self-consistency are included in the theory.  The pinning of the flat band to the Fermi energy is very sensitive to twist angle.  There is a sharp maximum close to $\theta = 1.05^\circ$ where the ultraflat conduction (and valence) band has a bandwidth of about 8 meV and persists for about 40 percent of the band filling.  Although decreasing the twist angle by just 0.1 degrees has a lower bandwidth of $\approx$ 2 meV, the band pinning persists only for $\approx$ 15 percent of the band filling.  This gives a new criteria for a ``magic window" range when the heavy fermion induced persistent flat band are most stable. 

\begin{figure}[th!]
\includegraphics[width= \linewidth]{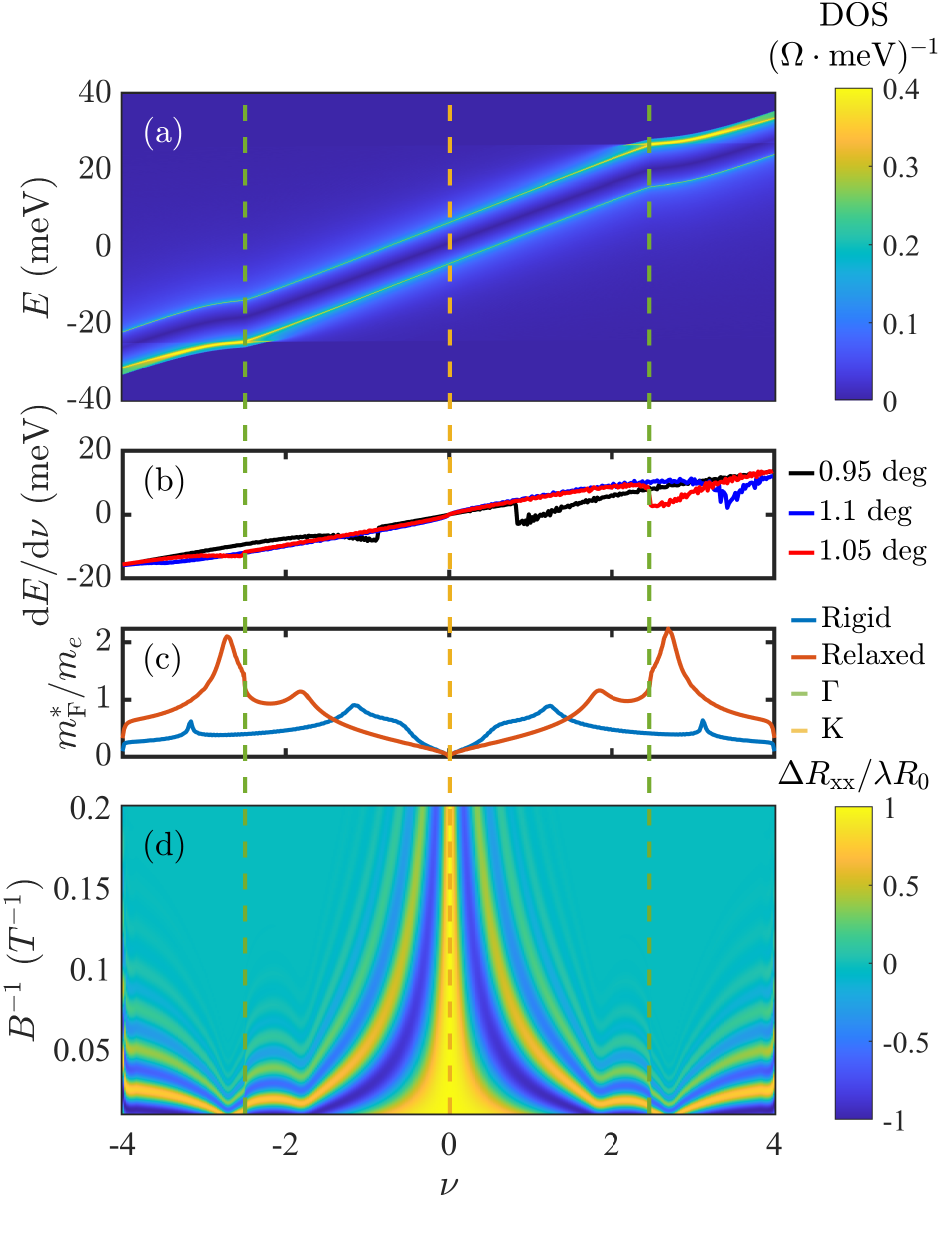}
\caption{Experimental signatures of the Lifshitz transition to the heavy fermion electron pocket and pinned ultraflat band.  The central orange dashed line marks charge neutrality and the green dashed line indicates $\nu_\Gamma$ for $\theta = 1.05^\circ$. (a) Map of density of states as a function of energy and filling. As understood from Eq.~\ref{Eq:perturbative}, the Van Hove singularity peaks first evolve linearly with filling for $|\nu| < |\nu_\Gamma|$.  After the Lifshitz transition at $\nu_\Gamma$, the Van Hove singularity is pinned in energy since the heavy fermion pocket allows for additional filling without incurring much Hartree cost. (b) Variation of total energy with filling.  The Lifshitz transition appears as sharp dips in $dE/d\nu$ occurring at $\pm \nu_\Gamma$. (c) Effective mass at the Fermi energy as a function of filling for the rigid (blue) and relaxed (red) models.  (d) Corresponding map of Shubnikov-de Haas oscillations as a function of inverse magnetic field and filling assuming a constant scattering time. The Lifshitz transition is visible as a sharp discontinuity.}
\label{Fig:EffectiveMass}
\vspace{-0.2in}
\end{figure}

Our predicted heavy fermion and ultraflat band can be probed experimentally.  Unfortunately since scanning probe spectroscopy measures the position of the Van Hove singularity relative to the Fermi energy, and since $E_{\rm vHS}$ and $E_F$ all have identical kinks, we expect the heavy fermion to be invisible to STM (see Supplemental Material).  However, as illustrated in Fig.~\ref{Fig:EffectiveMass}, we expect the heavy fermion pocket at the $\Gamma$-point should be observable in both compressibility experiments as a dip in $dE/d\nu$ and in Shubnikov-de Haas (SdH) transport oscillations.  The upper panel shows a color map for the density of states plotted as function of energy and filling.  At charge neutrality, the Hartree correction is zero and the two peaks in DOS correspond to the Van Hove singularities.  Close to $\nu =0$, one would observe the typical Dirac fermion signature. As one increases the filling, the Hartree energy is proportional to filling and this rigidly shifts the bands with constant slope. The kink at $\nu_\Gamma \approx 2.3$ and dip in $dE/d\nu$ can be seen in the figure.  Since the perfectly flat band remains unoccupied and transport measurements probe the Fermi energy, a kink is seen in the effective mass rather than a diverging electron mass.  We comment that Fig.~\ref{Fig:EffectiveMass}b looks very similar to the cascade physics reported in the literature~\cite{wong2020cascade, zondiner2020cascade, rozen_entropic_2021, saito_isospin_2021}.  However, we reiterate that the present work neglects the exchange splitting between spin and valley flavors. One obvious difference is that the heavy fermion dip we predict occurs at $\nu_\Gamma(\theta)$ that varies continuously as a function of $\theta$ within the ``magic window", while cascades are expected to occur only at integer fillings.  \\

\noindent \emph{Summary and Outlook}---  Although the heavy fermion in magic angle twisted bilayer graphene was anticipated previously \cite{song2022magic}, our theory is the first to provide the microscopic mechanism for its emergence.  We find that it arises from a purely mean-field electrostatic treatment provided both relaxation and self-consistency are included into the Hartree theory.  Since both the idea of Hartree corrections to the twisted bilayer graphene bands and lattice relaxation have been around for many years,   one might ask why the co-existing heavy fermion and ultraflat band was not anticipated previously.  Part of the reason is that relaxation was not fully realized in most previous treatments. A common approach is to obtain relaxed bands from density functional theory at higher twist angle and then assume that relaxation does not change as one approaches the magic angle.  Moreover, the shift of the magic angle with pseudomagnetic fields was only appreciated recently~\cite{ceferino2023pseudomagnetic,ezzi2023analytical}.  Given the narrow angle range over which the heavy fermion is stable, these partially relaxed models did not observe the Lifshitz transition.  Some theories focused on Lifshitz transitions within the non-interacting model~\cite{hejazi_multiple_2019, bennett2023twisted}.  While these reveal interesting physics, they are distinct from the Fermi surface topologies uncovered here.  

Many Hartree theories artificially used high values of dielectric constant to suppress the Hartree energy and get ``agreement" with experiment. Even though these authors may have calculated their Hartree interaction self-consistently, in actuality, they were still within the perturbative regime where no Lifshitz transition occurs.  Similarly, others just assumed that the Hartree energy was linear in filling thereby missing the kink completely.  Perhaps the closest to our work is a very recent GW-theory~\cite{zhu2024gw} where they also use self-consistency with a realistic dielectric constant and partially relaxed bands.  They also find a Lifshitz transition, but to a Fermi surface that is topologically distinct from ours.  Yet, like us, they also find an electron pocket at the $\Gamma$-point.  However, their $\nu^*$ is significantly less than $\nu_\Gamma$.  When their pocket at the $\Gamma$-point starts to fill, the M-point is much higher in energy.  As a result, they do not find the stabilization of a flat band by the heavy fermion that is our central result. 

Since our theory includes both full relaxation and self-consistency, it provides accurate bands for magic angle twisted bilayer graphene at high temperature.  This is not only important as a starting point to understanding the superconductivity \cite{cao2018unconventional,yankowitz2019tuning,lu2019superconductors}, correlated insulators \cite{cao2018correlated,lu2019superconductors}, and cascade physics \cite{wong2020cascade,zondiner2020cascade}, but more generally, because twisted bilayer graphene is the conceptual motif to understand the ever-increasing number of unexpected and unexplained phenomena in quantum two-dimensional materials \cite{liu2020tunable,xu2021tunable,park2022robust,kang2024evidence,lu2024fractional}. \\
 
\noindent \emph{\it Acknowledgements}--- This work was supported by the Singapore National Research Foundation Investigator Award (NRF-NRFI06-2020-0003).  We acknowledge Harshitra Mahalingam  for contributions during the early stages of this project.  It is a pleasure to thank Christophe De Beule and Eugene Mele for helpful comments on our manuscript.


\bibliographystyle{unsrt}
\bibliography{Bibliography}

\newpage 
\clearpage

\setcounter{equation}{0}
\setcounter{figure}{0}
\setcounter{table}{0}
\setcounter{page}{1}
\makeatletter
\renewcommand{\theequation}{S\arabic{equation}}
\renewcommand{\thefigure}{S\arabic{figure}}
\renewcommand{\bibnumfmt}[1]{[S#1]}
\renewcommand*{\thepage}{A\arabic{page}}
\renewcommand{\thesection}{\Alph{section}}
\renewcommand{\thesubsection}{\arabic{subsection}}

\setcounter{secnumdepth}{2}
\twocolumngrid

\begin{widetext}
    \begin{center}
    \textbf{\large Supplemental Materials for: A self-consistent Hartree theory for lattice-relaxed magic-angle twisted bilayer graphene}
\end{center}
\part{} 
\parttoc 
\end{widetext}

\section{\label{sm_mp}Choice of Model Parameters}

As discussed in the main text, we use six parameters in the model.  Here we discuss each of them in more detail explaining our choices.  The graphene monolayer Fermi velocity and lattice constant $a_0 = 0.246~\rm{nm}$ are established in the literature.  Our value of $v_{\rm F}= 1.05 \times 10^{6}~{\rm m/s}$ has been measured in numerous experiments over the past decade \cite{sarma2011electronic} and as shown in Fig.~\ref{FIG:Validation} agrees with STM measurements away from magic angle.  This value is also understood from theoretical calculations once the Fock contributions of the charge-neutral monolayer Dirac cones are included in a realistic calculation \cite{tang2018role}.

\begin{figure*}[ht!]
\includegraphics[width= 0.9\linewidth]{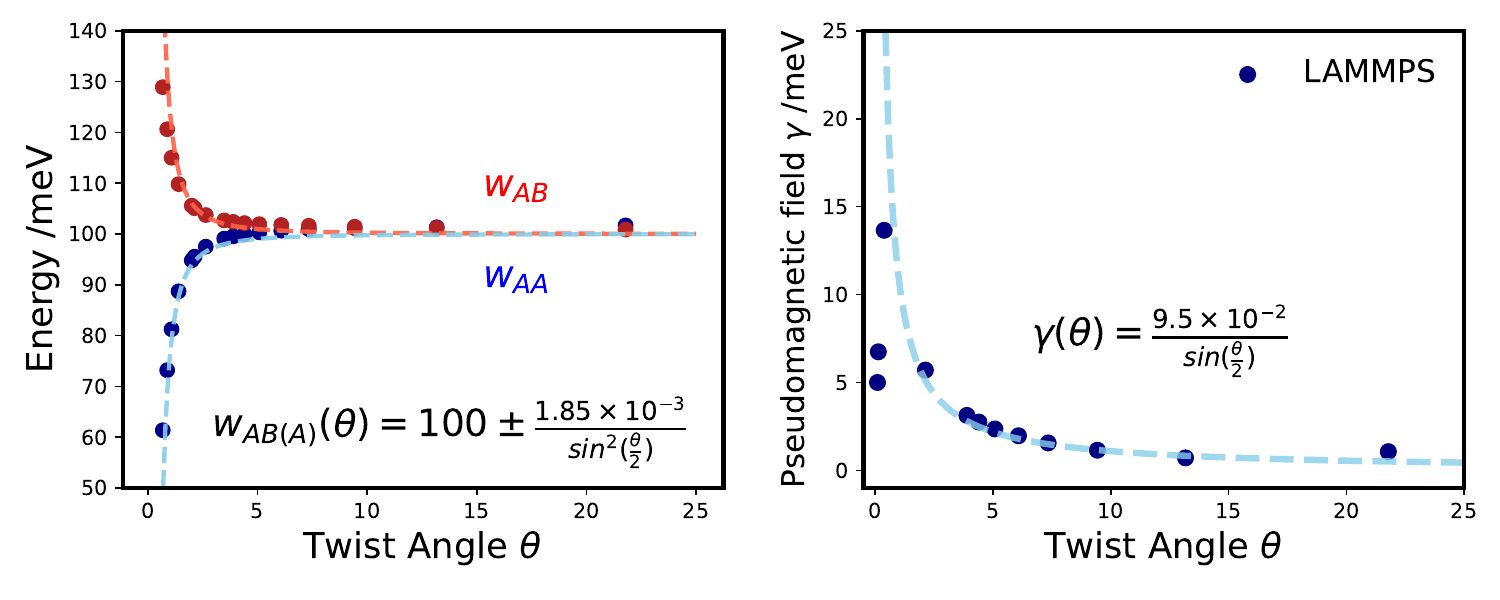}
\caption{Effect of atomic relaxation on the electronic continuum model parameters. Left: Interlayer coupling parameters linking electronic states between the two layers. The red line represents the same-sublattice coupling $W_{AA}$, while the blue line is for different-sublattice coupling $W_{AB}$. Points denote calculated data, and dashed lines indicate fitted values based on the provided fitting function. For large twist angles, atomic relaxation diminishes rendering both coupling parameters identical. Conversely, for small twist angles atomic relaxation becomes significant favoring AB/BA regions over AA regions. This asymmetry results in an increase in opposite-sublattice coupling, which is inversely correlated with the decrease in same-sublattice coupling. Right: Pseudomagnetic field strength as a function of twist angle. Data points are represented by dots and the fit by a dashed line. The fitting accurately describes the pseudomagnetic strength until approximately 0.5 degrees, beyond which discrepancies arise. The emergence of a pseudomagnetic field arises from atomic relaxation-induced alterations in bond lengths at the monolayer level.}
\label{Fig:Relaxation}
\end{figure*}

The interlayer \moire potential at large twist angles where lattice relaxation becomes unimportant is taken to be $W^{*} = 100~{\rm meV}$.  In this rigid case atoms do not move, leading to two effects: First, the matrix elements connecting all sublattice elements are the same because, on average, they have the same density. Second, the atomic arrangement of atoms varies smoothly and decays very slowly as a function of spatial distance. This only leads to first harmonic tunneling, and higher-order tunneling dies very quickly.  In this case, $W^{*} = W_{AA}(\theta \gg 1) = W_{AB}(\theta \gg 1) = f~t_\perp$ is a fraction of the Bernal bilayer graphene interlayer tunneling potential, and $f\approx 0.4$~\cite{dos2007graphene}.  Here $W_{AA}$ and $W_{AB}$ are the \moire interlayer hopping elements between sublattice A in one layer and either sublattice $A$ or $B$ in the second layer.   Following Ref.\cite{ezzi2023analytical}, we obtain $W^*$ by first determining the relaxed atomic positions using LAMMPS molecular dynamics simulations, then using the standard Slater-Koster parameterization, determine the Fourier components of the \moire potential connecting electronic states from both layers.  This is illustrated in Fig.~\ref{Fig:Relaxation}. 


For small twist angles, atoms in twisted bilayer graphene (TBG) change their nominal monolayer graphene positions to minimize energy. The geometry of TBG encompasses all possible sorts of vertical stacking \cite{dos2012continuum}, which do not have the same total energy \cite{alden2013strain}. The vertical van der Waals interaction drive atoms into favorable vertical stacking, while the elastic force between atoms in the panel tends to maintain the hexagonal structure of monolayer graphene. The physical mechanism for atomic relaxation is to increase the regions with minimum energy (AB/BA regions) while minimizing the size of high energy stackings (AA regions). There are many approaches to atomic relaxation \cite{nam2017lattice,carr2019exact,guinea2019continuum,ezzi2023analytical,pan2024structural} which mostly provide consistent results, thereby supporting physical insights on the  nature of atomic relaxation \cite{ezzi2023analytical}.  Atomic relaxation affects inter-layer electronic parameters as well as intra-layer electronic parameters. With asymmetric sizes of high symmetry regions, 
in contrast to the rigid model the hopping between same sublattice  orbitals (AA and BB) and different sublattice (AB and BA) are not the same.  The main effect of lattice relaxation is that $W_{AA}$ and $W_{AB}$ are no longer equal.  We define  $W_{AB(A)}(\theta)= W^* \pm  \Delta W(\theta)$. We find $\Delta W(\theta) = A/\sin^2(\theta/2)$, with $A = 1.85 \times 10^{-3}~{\rm meV}$.

As shown in Fig.~\ref{Fig:Relaxation}, we obtain $\Delta W(\theta)$ by fitting to LAMMPS data \cite{ezzi2023analytical}.  However, we note that the order of magnitude of this effect can also be estimated as $A \approx (1/8 \pi) (V_{\rm vdW}/(\mu + \lambda)) V^0_{pp \sigma}$, where $V_{\rm vdW}$ is the van der Waals potential between the two layers, $\mu$ and $\lambda$ are graphene Lam\'e coefficients and  $V^0_{pp \sigma}$ is the Slater-Koster parameter for vertically oriented carbon $\pi$-orbitals.

Relaxation of atoms in TBG also locally change the bond lengths in monolayer graphene and therefore breaking the $C_3$ symmetry at the monolayer level. This  shifts the position of the Dirac cones, leading to the emergence of a gauge field that couples to the Dirac electrons when the tip of the cone is considered at original Dirac point. This leads to what is known as a pseudomagnetic field.  Also shown in the right panel of Fig.~\ref{Fig:Relaxation} the pseudomagnetic vector potential $\gamma(\theta) = B /\sin(\theta/2)$, where $B =9.5 \times 10^{-2}~{\rm meV}$ \cite{ezzi2023analytical,ceferino2023pseudomagnetic}.  We can also estimate $B \approx 4 \pi \beta_G (V_{\rm vdW}/(\mu + \lambda))$ where $\beta_G$ is the Gr\"uneisen parameter \cite{mohiuddin2009uniaxial}.

\begin{figure*}[ht!]
\includegraphics[width= \linewidth]{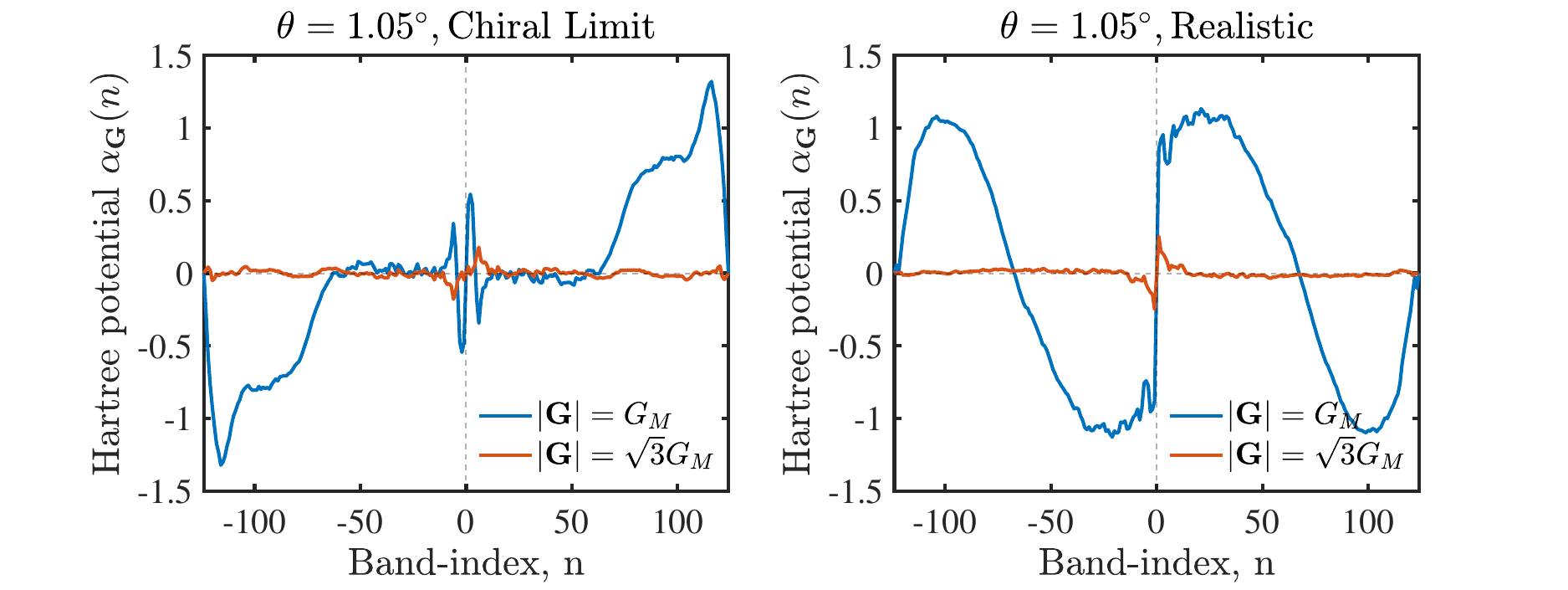}
\caption{Oscillation of the total Hartree potential with chemical potential. In both panels, the total Hartree potential coefficient is plotted on the y-axis against the number of bands summed from the bottom up to a given band index on the x-axis. The blue trace represents the expansion coefficient of the first shell associated with the magnitude of the reciprocal lattice vector $G_M$, while the brown trace is for the second shell. Left: results for the chiral limit model. Right: results for the realistic model. In both cases and for both shells, oscillations in the Hartree potential are evident. These oscillations result in a vanishing Hartree potential when summed up to charge neutrality, a critical outcome for the equivalence of the two subtraction schemes discussed in this study.}
\label{oscillations1}
\end{figure*}

As emphasised in the main text we take the substrate dielectric constant $\epsilon =4$.   This sets the strength of the electron-electron interactions.  Typical substrates like SiO$_2$ and hBN have dielectric constant values close to $\epsilon = 4$ (and there is some variation depending on whether the sample is fully encapsulated as is typical in transport, or with one surface exposed as is typical in STM).  Typically the Coulomb interaction $V(q, \epsilon, d)$ also includes the distance $d$ from the metallic screening gate.  We take $d \rightarrow \infty$ and show in Sec.~\ref{Sec:Benchmarking} that the precise value is unimportant for $d> 2~{\rm nm}$.  In that section we also show that although there are other auxiliary parameters in our code such as the number of bands we include in the calculation, and the number of shells of the basis functions etc., we have carefully calibrated our codes so that these parameters are chosen to be large enough that our numerics converge.

\section{Hartree Interaction Subtraction Schemes and their Equivalence within the One-shot Hartree Approximation}

To calculate the effect of the Hartree potential on the low-energy bands of twisted bilayer graphene, one needs to first decide which charges are going to contribute to the Hartree potential. In the literature, this is done in basically two different ways. The first scheme considers the electronic contribution of all states from the bottom of the Fermi sea until the chemical potential, while the second considers only the states between the charge neutrality point and the chemical potential.  The purported reason for these different approaches is to avoid double counting the Coulomb interaction both before and after including the \moire potential.  In this section, we show analytically and numerically the conditions for the equivalence between the various schemes and demonstrate when and why they disagree.

The Hartree potential is periodic and can be expanded in a Fourier sum as: 

\begin{equation}
\rho(r)=\sum_{G}\alpha_{G}e^{ir\cdot G}
\end{equation}

\noindent with the expansion coefficients: 
\begin{equation}
\alpha_{G}=\sum_{n,k,\alpha,G^{\prime}}\phi_{G^{\prime}}^{*\alpha,n}(k)\phi_{G+G^{\prime}}^{\alpha,n}(k)
\label{eq:coeff_HTpotential}
\end{equation}

\noindent where $\phi$ stands for wavefunctions,  $n$ is the band index, $k$ is k-point index, $\alpha$ is sublattice index and $G^{\prime}$ is reciprocal lattice vector index. This can be rewritten as $\alpha_{G} = \sum_{k} \sum_{n} \chi_{n,k}$, where

\begin{equation}
\chi_{n,k}= \sum_{\alpha,G^{\prime}}  \phi_{G^{\prime}}^{* \alpha,n}(k)\phi_{G+G^{\prime}}^{\alpha,n}(k) 
\label{gamma}
\end{equation}

\begin{figure}[th!]
\includegraphics[width= 1 \linewidth]{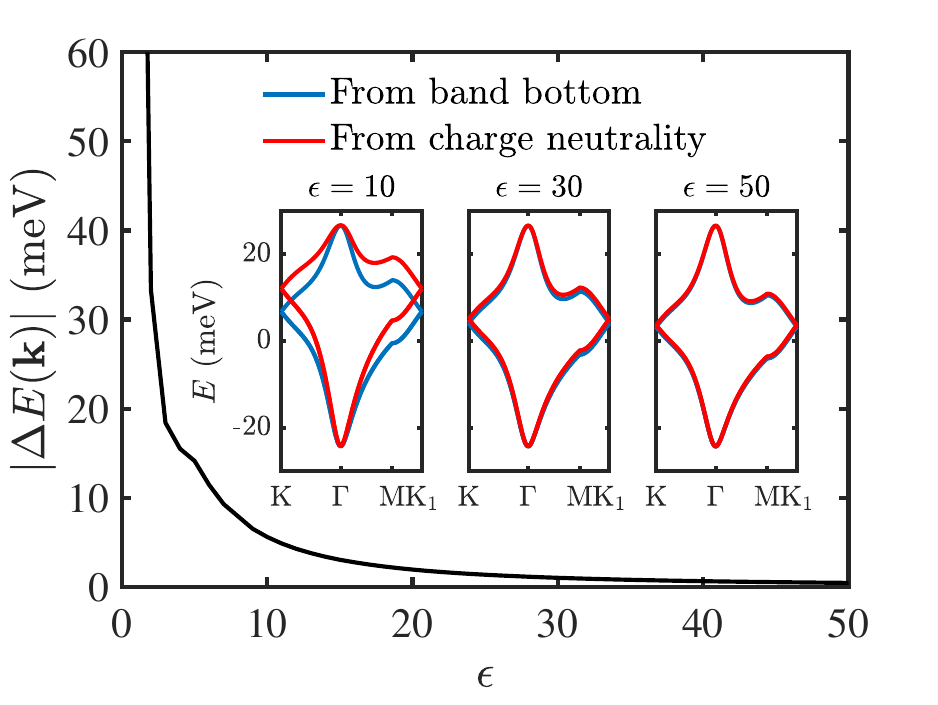}
\caption{Numerical difference between the two subtraction schemes. Main: The maximum energy difference, for the two low-energy bands, between the two subtraction schemes as a function of the dielectric constant $\epsilon$. As $\epsilon$ increases, the difference between the two schemes decreases. The dielectric constant $\epsilon$ encodes information about the asymmetry between particle-hole spectra. Inset: Band structures for three representative values of $\epsilon$ for both schemes. Scheme 1 (light blue) sums from the bottom of the band until the chemical potential, while Scheme 2 (red) sums from the charge neutrality point to the chemical potential. For small values of $\epsilon$, particle-hole asymmetry is dramatic, resulting in a high difference (in the main panel). Conversely, for large values of $\epsilon$, the asymmetry diminishes, leading to a identical results between the two subtraction schemes.}
\label{Fig:SubtractionSchemes}
\end{figure}

\noindent To gain intuition into this problem, we start by looking first at the chiral model of twisted bilayer graphene \cite{san2012non,tarnopolsky2019origin}.  The chiral model is a sublattice-symmetric model (like monolayer graphene) with the following schematic representation of the Hamiltonian:

\begin{equation}
   H= \begin{pmatrix}
        0 & H_{AB} \\
        H_{AB}^\dagger & 0
    \end{pmatrix}
\end{equation}

\noindent The Hamiltonian is written with basis divided into two groups, the first group contains the wavefunctions of sublattice A (regardless of their layer index), and the second group consists of wavefunctions with sublattice index B.

\noindent A defining property of a sublattice symmetric Hamiltonian is:

\begin{equation}
   \sigma_z H \sigma_z = - H 
   \label{sublttice}
\end{equation}

\noindent $\sigma_z$ is a diagonal matrix that equals $+1$ for sublattice A and $-1$ for sublattice B.  Sublattice symmetry gives rise to a symmetric spectrum and related wavefunctions. As can be seen from Eq.~\ref{sublttice}, if $\Psi_{n}= (\psi_A,\psi_B)$ is an eigenvector of the Hamiltonian with energy $\epsilon$ (above the charge neutrality point), then $\Psi_{-n}=(\psi_A,-\psi_B)$ is an eigenvector with energy $\epsilon_{-n} =- \epsilon$ (below the charge neutrality point). This relationship in the chiral model allows us to make analytical statements about the nature of the Hartree interaction. 

The Hartree coefficients (Eq.~\ref{gamma}) are sums of bilinear terms with the same sublattice index $\alpha$, leading to  $\chi_{-n,k}= \chi_{n,k}$.  For any Hamiltonian, filling all the bands has zero potential fluctuations implying that the Hartree contribution from all the bands sums up to zero: 

\begin{equation}
  \bar{\alpha}_{G} = \sum_{k} \sum_{n=1}^{N_{\text{bands}}} \chi_{n,k} = 0  
  \label{alpha}
\end{equation}

 \noindent which is a consequence of the completeness relation for each $k$ point. Writing this for one k-point in Eq.~\ref{alpha}, we have:
\begin{equation*}
 \chi_{-\frac{N}{2},k}+ ... +\chi_{-2,k}+ \chi_{-1,k} + \chi_{1,k} + \chi_{2,k}+ ... + \chi_{\frac{N}{2},k} =0
\end{equation*}

\noindent where $N$ is the total number of bands and $-1$ and $1$ label the valence and conduction bands, respectively.

Using $\chi_{-n,k}= \chi_{n,k}$, we have  
 $2(\chi_{-\frac{N}{2},k}+ ... +\chi_{-2,k}+ \chi_{-1,k} ) =0$ implying that $\chi_{-\frac{N}{2},k}+ ... +\chi_{-2,k}+ \chi_{-1,k}  =0$. The left-hand side is nothing but the expansion coefficient of the Hartree potential summed over the bottom of the bands until charge neutrality:

\begin{equation}
        \bar{\bar{\alpha}}_{G} = \sum_{k} \sum_{n=1}^{N_{\text{CNP}}} \chi_{n,k} = 0. \quad 
        \label{CNP}
\end{equation}

Equation \ref{CNP} shows that the Hartree contribution from all the bands below charge neutrality point sums up to zero, a crucial result which we use below to show the equivalence of the two subtraction schemes.

The first subtraction scheme sums the contributions of all occupied states from the lowest energy band until the chemical potential \cite{xie2020nature}.  In this case, the expansion coefficients can be decomposed as

\begin{equation}
    \begin{split}
        \alpha_{G} &= \sum_{k} \left(\sum_{n=1}^{\text{CNP}} \chi_{n,k} + \sum_{n=\text{CNP}+1}^{x} \chi_{n,k}\right) 
    \end{split}
    \label{Mac}
\end{equation}

The alternative subtraction scheme only sums contributions from the charge neutrality point to the chemical potential \cite{cea2019electronic}.  In this case, the expansion coefficients are

\begin{table}[t]
    \centering
    \caption{Table of energy slopes with filling at the K and $\Gamma$ points calculated using~\eqnref{oneshotslope} and numerically. All values are in meV and calculated at $\theta=1.05^\circ$. The analytical expression has excellent agreement with one-shot Hartree while the self-consistent values are suppressed, also seen in~\figref{os_vs_sc}.}
     \begin{tabular}{|c|c|c|}
        \hline
         & K-point& $\Gamma$-point \\
        \hline
        Semi-analytic (Eq.~\ref{oneshotslope}) & 25 {\rm meV} & 1 {\rm meV}\\
        \hline
        Relaxed, One-shot (numerics) & 25 {\rm meV} & 0.3  {\rm meV}\\
        \hline
        Relaxed, Self-consistent (numerics) & 8  {\rm meV}& 0.1 {\rm meV} \\
        \hline
        Rigid, One-shot (numerics) & 25 {\rm meV} & -0.9  {\rm meV}\\
        \hline
        Rigid, Self-consistent (numerics) & 7  {\rm meV}& -0.2 {\rm meV} \\
        \hline
    \end{tabular}
    \label{tab:pert_slopes}
\end{table}

\begin{equation}
    \begin{split}
        \alpha_{G} &= \sum_{k} \sum_{n=\text{CNP}+1}^{x} \chi_{n,k}
    \end{split}
    \label{Paco}
\end{equation}

\noindent Because of Eq.~\ref{CNP}, the first term in Eq.~\ref{Mac} is zero and we end up with the two methods giving identical results for the chiral model.  The left panel of Fig.~\ref{oscillations1} illustrates how the Hartree contribution fluctuates with changes in the chemical potential, beginning from the lowest band. The total Hartree contribution vanishes as the chemical potential reaches the charge neutrality point, and again when all the bands are completely filled.  This double oscillation property of the Hartree potential as a function of band number explains why the two subtraction schemes are equivalent for the single-shot Hartree interaction within the chiral model.  In the right panel of Fig.~\ref{oscillations1} we show numerically that the double oscillation persists for the realistic model thereby establishing that the full Hamiltonian also exhibits the same property.

This equivalence breaks down beyond the single-shot Hartree interaction. 
 As understood above, the equivalence deteriorates in proportion to the asymmetry between the particle and hole components of the spectrum.  It is this symmetry in the non-interacting Hamiltonian that guarantees that equivalence of the two substraction schemes.  However, since the Hartree interaction itself breaks this particle-hole symmetry (see Fig.~\ref{Fig:SubtractionSchemes}), the self-consistent Hartree results will be different in the two schemes unless the dielectric constant $\epsilon$ is large.  For realistic values of $\epsilon$ the differences between the two schemes are comparable or even larger than the bandwidth.  In the next section, we understand using perturbation theory how these differences arise.

 \begin{figure}[th!]
\includegraphics[width=1\linewidth]{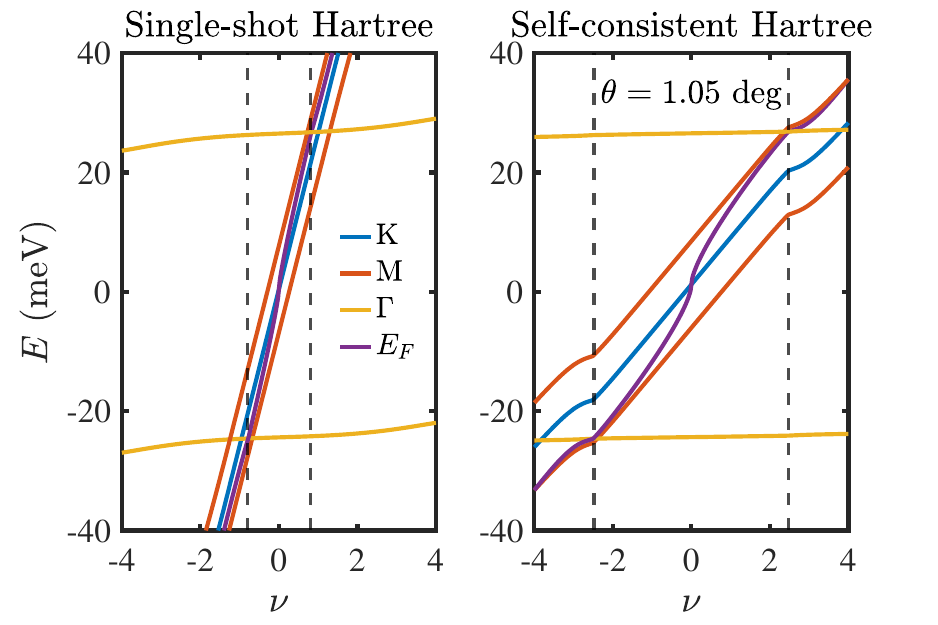}
\caption{ Evolution of band energies at the K (blue), M (red) and $\Gamma$ (yellow) points, and of the Fermi energy (purple) as a function of filling with one-shot (left) Hartree and self-consistent (right). In both cases, the band energies are seen to vary linearly with filling over a large range, justifying a perturbative analysis. Self-consistency is seen to reduce the sensitivity of the energy to filling, and introduce kinks in the band energies where the M and $\Gamma$ energies coincide, that is where the band is heuristically flattest.}
\label{os_vs_sc}
\end{figure}

\begin{figure}[th!]
\includegraphics[width=1\linewidth]{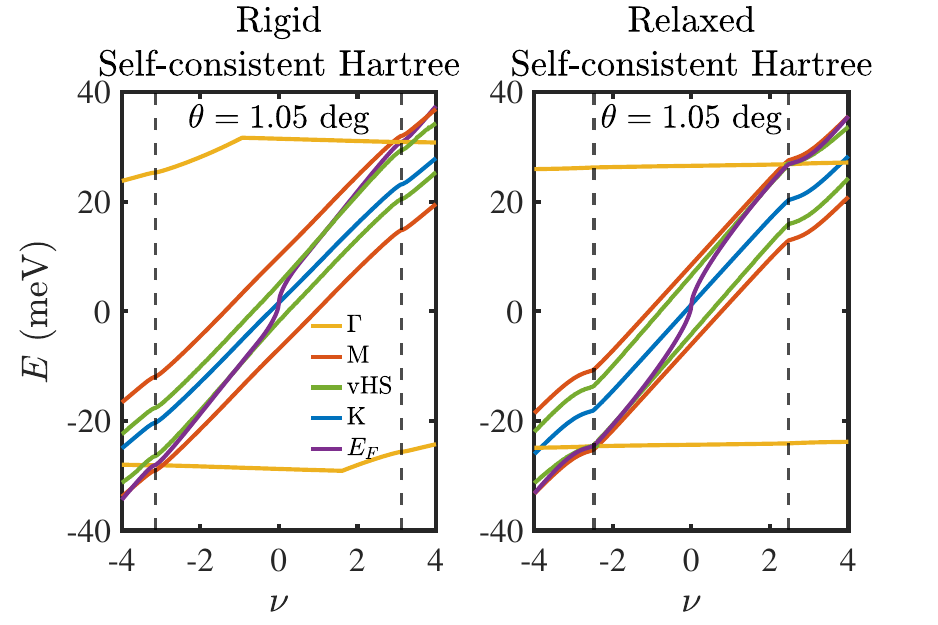}
\caption{Energy shifts as a function of filling for rigid and relaxed model.  The Fermi energy (purple) is shown along with the energies of the $\Gamma$, $M$, and $K$ points as well as the Van Hove singularity.  Notice that the relaxed model has $\nu_\Gamma = E_\Gamma = E_M = E_{\rm VHS}$.}
\end{figure}

\begin{figure}[h]
\includegraphics[width=1\linewidth]{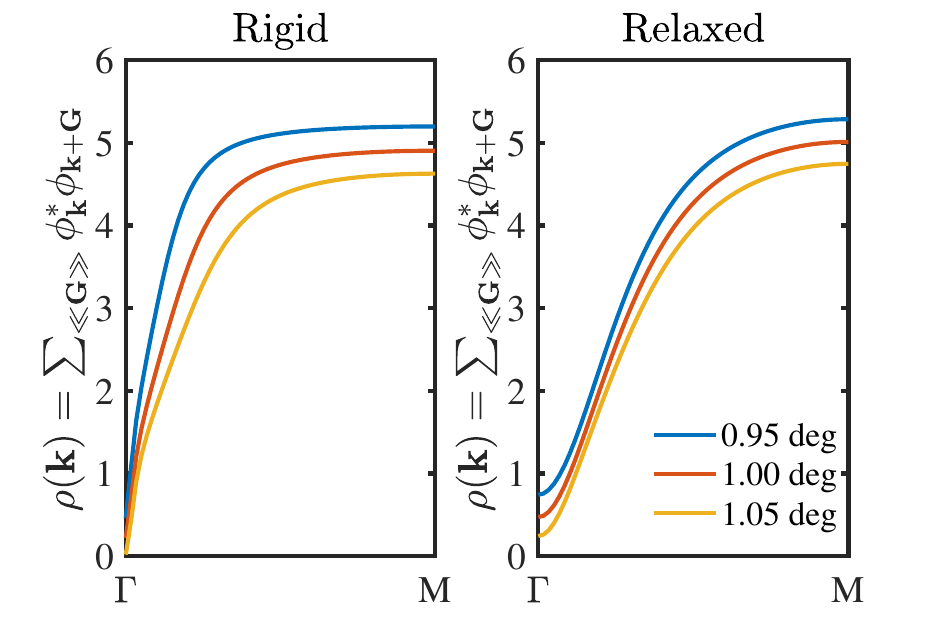}
\caption{Effect of relaxation on form factor.  Cuts along the high symmetry $\Gamma$ to $M$ path for different twist angles.  Relaxation smooths the variance in form factor along this path.}
\label{Fig:FormFactor}
\end{figure}

\section{Perturbative Analysis of the Hartree Term}

The full treatment of the Hartree term requires self-consistency, making it difficult to extract the essential features analytically. However, we note that for a wide range of filling factors centered at charge neutrality, the band energies shift linearly with filling, giving legitimacy to a perturbative analysis with perhaps renormalized scale factors.  We start by  discussing the one-shot Hartee correction.  In the basis in which the non-interacting Hamiltonian is diagonal we have  
 \begin{align}  H_\mathrm{Har}=4\sum_{nn'}\sum_{\bfk}\sum_{\bfG\neq0} \left( U_\bfG\rho_\bfG\Lambda^{nn'}_{\bfk,-\bfG} \right) \psi^\dagger_{n,\bfk}\psi_{n',\bfk} \label{sm_hardef}
\end{align}
$n$ and $n'$ are band indices.  Here, the density harmonics are given by 

\begin{align}    \rho_\bfG&=\sum_{n,\bfk}\Lambda^{nn}_{\bfk,\bfG}\langle\psi^\dagger_{n,\bfk}\psi_{n,\bfk}\rangle\\
    &\approx\frac{\nu}{4}\Lambda^{\mathrm{sgn}(\nu)}_{\mathrm{K},\bfG},\label{os_rhog}
\end{align}

\noindent where in the last line we use $\Lambda^{nn}=\Lambda^{n}$ to simplify notation and we begin to introduce assumptions pertinent to a perturbative approach. First, we evaluate the expectation values with respect to the noninteracting bands. This means that at charge neutrality, $\rho^{\mathrm{CN}}_\bfG=0$, as shown in the previous section.  We can therefore restrict the sum to deviations from charge neutrality. Furthermore, for a perturbative analysis, we only ever fill the upper central band or deplete the lower central band, which we denote with $+$ and $-$ respectively, with $\mathrm{sgn}(\nu)\in\{+,-\}$, completely removing the sum over $n$. Finally, we note that electrons are initially added to the K and K$'$ points, allowing us to approximate $\Lambda_\bfk$ by the constant $\Lambda_\mathrm{\bfk=K}$ and pull it out of the sum over $\bfk$. The remaining sum over occupied states is the filling factor $\nu$.

The form factors result in a Hartree correction that is not diagonal in the noninteracting basis. However, taking the filling $\nu$ as a small parameter, these interband corrections are $\mathcal{O}(\nu^2)$ and so we focus on the intraband contributions, giving to first order in $\nu$,

\begin{align}
    \delta\varepsilon_{n,\bfk}(\theta,\nu)&=\frac{2\pi e^2}{\epsilon A_m}\nu\sum_{\bfG\neq0}\frac{\Lambda^{\mathrm{sgn}(\nu)}_{\mathrm{K},\bfG}\Lambda^n_{\bfk,-\bfG}}{|\bfG|}\\
    &\approx\frac{6e^2\theta}{\epsilon a}\nu\Lambda^{\mathrm{sgn}(\nu)}_{\mathrm{K},\bfG_0}\Lambda^{n}_{\bfk,-\bfG_0}, \label{oneshotslope}
\end{align}
where $\delta\varepsilon$ is the shift between the interacting and noninteracting bands, and in the second line we restrict the sum over $\bfG$ to the first star, with $\bfG_0$ a characteristic vector of that star and the factor of 6 accounting for the symmetry equivalent vectors in this first star. This form allows us to separate the contributing elements as the characteristic energy scale $\frac{e^2\theta}{\epsilon a}\approx25.5$ meV at $\epsilon=4$ and $\theta=1^\circ$ which acts between \moire cells, the filling factor $\nu$ and form factors corresponding to the band generating the Hartree shifts, $\Lambda^{\mathrm{sgn}(\nu)}$, and the band getting shifted, $\Lambda^n$. In Fig.~\ref{Fig:FormFactor} we show the effect of relaxation on the form factors.

Table~\ref{tab:pert_slopes} summarises the linear slopes of $\delta\varepsilon_{n,\bfk}(\theta,\nu)$ against $\nu$, showing excellent agreement between the perturbative analytic treatment and one-shot Hartree, which assumes the wavefunctions are not appreciably altered by the interaction. The self-consistently calculated values meanwhile are systematically lower than those calculated in one-shot. As seen in~\figref{os_vs_sc} it remains true that the energy shifts are indeed linear in $\nu$ for a wide range of $\nu$ even with self-consistency.  This inspires us to extend the purturbative analysis to the self-consistent case.

We note that the sole difference between one-shot and self-consistent Hartree lies in the evaluation of the density harmonics, 

\begin{align}
    \tilde{\rho}_\bfG&=\sum_{n,\bfk}\tilde{\Lambda}^{nn}_{\bfk,\bfG}\langle\tilde{\psi}^\dagger_{n,\bfk}\tilde{\psi}_{n,\bfk}\rangle\\
    &\approx\tilde{\rho}^{\mathrm{CN}}_\bfG+\frac{\nu}{4}\tilde{\Lambda}^{\mathrm{sgn}(\nu)}_{\mathrm{K},\bfG},\label{sc_rhog}
\end{align}
where the tilde denotes the self-consistent band basis instead of the noninteracting band basis used in one-shot Hartree. Previously we had $\rho^{\mathrm{CN}}_\bfG=0$, but as discussed in the previous section, this is no longer true when using the self-consistent bands and wavefunctions. We note however that self-consistency does not change the fact that the bands begin filling from the K points, justifying the second term of~\eqnref{sc_rhog} which is, at this point, the only approximation made so far. The first term may be evaluated as

\begin{align}
    \tilde{\rho}^{\mathrm{CN}}_\bfG&=\sum_\bfk\sum_{n=-\infty}^{\mathrm{CN}}\tilde{\Lambda}^n_{\bfk,\bfG}\langle\tilde{\psi}^\dagger_{n,\bfk}\tilde{\psi}_{n,\bfk}\rangle\\
    &\approx\sum_\bfk\left(\tilde{\Lambda}^-_{\bfk,\bfG}-\Lambda^-_{\bfk,\bfG}\right),\label{sc_dens}
\end{align}
where in the second line we make use of two assumptions. First, if the conduction and valence bands remain separated except at K, that is if there is no energy level which crosses both bands, then the occupation of all states for bands below charge neutrality is unity, removing the expectation value. This assumption is borne out in our numerical calculations. Second, owing to the large band gap introduced by relaxation, the wavefunctions and energies of all bands below the valence band are essentially unchanged, $\tilde{\Lambda}^n=\Lambda^n$ for $n$ below the valence band.

We then subtract $\rho^{\mathrm{CN}}_\bfG=\sum_\bfk\sum_{n=-\infty}^{\mathrm{CN}}\Lambda^n_{\bfk,\bfG}=0$ to obtain~\eqnref{sc_dens}, which retains the feature that if the terms with a tilde are evaluated using the noninteracting band basis, the charge neutral density harmonics vanish.  The problem now reduces to finding the self-consistent form factor in terms of the noninteracting form factors and the filling.  In line with our assumption that the distant bands are unmodified by interaction, we restrict our attention to the conducting and valence bands. The Hamiltonian in this 2-band basis may be written as

\begin{figure*}[t!]
\includegraphics[width= 0.95\linewidth]{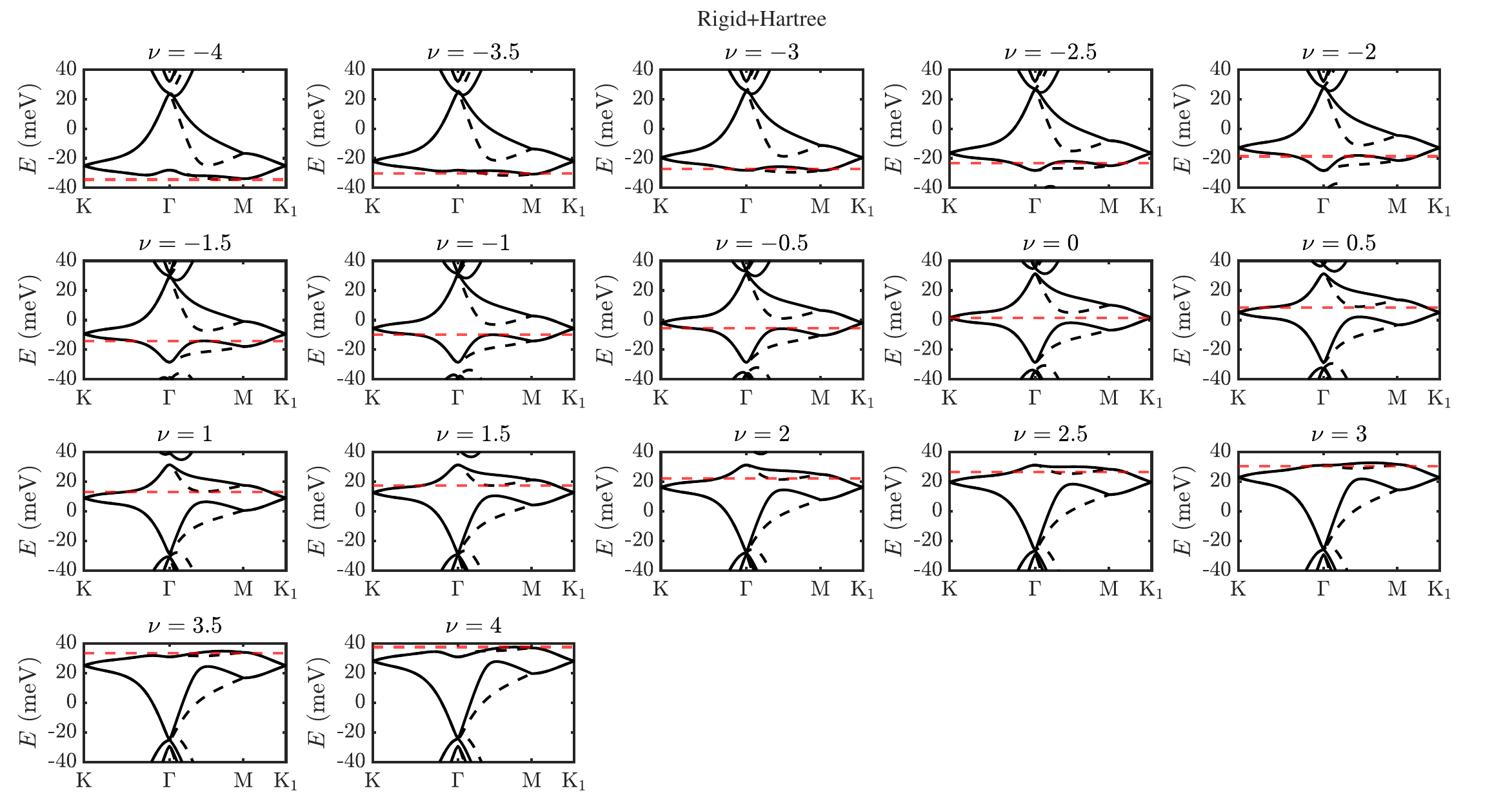}
\caption{Valence and conducting bandstructures for the rigid model with self-consistent Hartree interactions at $\theta=1.05$ deg, from filling factor $\nu=-4$ to $\nu=4$. At $\nu=-4$, the valence band is gapped from lower energy bands, while the conducting band crosses higher energy bands. Valence asymmetry remains visible in both bands at all fillings. Filling starts at the M point and progresses smoothly to the $\Gamma$ point at $\nu\approx-3$ and the K points at $\nu=0$, with the filling order and gap structure reversing for $\nu>0$. The curvature at the $\Gamma$ point in the valence (conduction) band reverses sign at $\nu\approx-3.5$ ($\nu\approx3.5$) before (after) the $\Gamma$ point is filled.}
\label{Fig:Bands_with_EF_rigid_Hartree}
\end{figure*}

\begin{align}
    H&=\sum_\bfk\tilde{\Psi}^\dagger_\bfk\tilde{\varepsilon}_\bfk\tilde{\Psi}_\bfk\\
    &=\sum_\bfk\Psi^\dagger_\bfk\left(\varepsilon_\bfk+\tilde{V}_\bfk\right)\Psi_\bfk,
\end{align}
where $\Psi^\dagger=(\psi^\dagger_+,\psi^\dagger_{-})$, $\varepsilon$ and $\tilde{\varepsilon}$ are diagonal matrices whose entries are the noninteracting and self-consistent band energies respectively and $\tilde{V}_\bfk$ is a matrix whose elements are the self-consistently evaluated Hartree energy,

\begin{align}
    \tilde{V}^{nn'}_\bfk=4\sum_{\bfG\neq0}\tilde{\rho}_\bfG U_\bfG\Lambda^{nn'}_{\bfk,-\bfG},\label{V_defn}
\end{align}
as found in~\eqnref{sm_hardef}. Diagonalization is achieved through a unitary transformation, which gives both the self-consistent wavefunctions and the Bloch coefficients as

\begin{align}
    \tilde{\Psi}_\bfk&=\mathcal{U}_\bfk\Psi_\bfk\\
    (\tilde{c}_{+,\bfk+\bfG},\tilde{c}_{-,\bfk+\bfG})&=(c_{+,\bfk+\bfG},c_{-,\bfk+\bfG})\mathcal{U}^{-1}_\bfk,
\end{align}
with the transformation explicitly given as

\begin{align}
    \mathcal{U}&=
    \begin{pmatrix}
        w & z \\
        -z^* & w
    \end{pmatrix}\\
    w&=\sqrt{1-|z|^2}\label{w_defn}\\
    z&=\alpha \tilde{V}^{+-}\label{z_defn}\\
    \alpha^2&=\frac{2}{\Delta\varepsilon^{\prime 2}+4|\tilde{V}^{+-}|^2+\Delta\varepsilon'\sqrt{\Delta\varepsilon^{\prime 2}+4|\tilde{V}^{+-}|^2}}\label{alpha_defn}\\
    \Delta\varepsilon'&=\varepsilon^+ + \tilde{V}^{++}-\varepsilon^- -\tilde{V}^{--}\label{deleps_defn},
\end{align}
and we suppress the $\bfk$ index common to all quantities. The diagonal components of the form factor then transform as

\begin{align}
    \tilde{\Lambda}^\pm=w^2\Lambda^\pm+|z|^2\Lambda^\mp\pm w\left(z\Lambda^{-+}+z^*\Lambda^{+-}\right),\label{sc_ff}
\end{align}
where we suppress the $\bfk$ and $\bfG$ indices. In this language, the self-consistency manifests in that the $\mathcal{U}$ that transforms from the noninteracting to interacting diagonal basis itself depends on the $V$ and thus quantities evaluated in the interacting basis.

Recalling our numerical results and how they suggest a perturbative analysis, we expand in the filling $\nu$ noting that $\tilde{\rho}^\mathrm{CN}_\bfG$ vanishes at $\nu$=0, and thus, to lowest order, $\tilde{\rho}^\mathrm{CN}_\bfG\approx\frac{\nu}{4}\delta\Lambda_{\bfG}$ which, in conjunction with Eqns.~\ref{sc_dens},~\ref{V_defn} and~\ref{w_defn} through to~\ref{sc_ff}, give

\begin{align}
    \delta\Lambda_{\bfG_0}\!=\!-\frac{48e^2\theta}{\epsilon a}\!\left(\sum_\bfk\frac{\Lambda^{+-}_{\bfk,-\bfG_0}\Lambda^{-+}_{\bfk,\bfG_0}}{\Delta\varepsilon_\bfk}\!\right)\!\left(\!\Lambda^{\mathrm{sgn}(\nu)}_{\mathrm{K},\bfG_0}\!+\!\delta\Lambda_{\bfG_0}\!\right),
\end{align}
as the defining equation for the density fluctuation at charge neutrality with $\Delta\varepsilon=\varepsilon^+-\varepsilon^-$. The factor of 48 comes from the sixfold symmetry of the first star, a factor of 4 in the definition of $\nu$ and a factor of 2 from the linear in $z$ parts of~\eqnref{sc_ff}, which give the contributions linear in $\nu$.

This ultimately leads to an energy shift of

\begin{align}
    \delta\varepsilon_{n,\bfk}(\theta,\nu)&=\eta\frac{6e^2\theta}{\epsilon a}\nu\Lambda^{\mathrm{sgn}(\nu)}_{\mathrm{K},\bfG_0}\Lambda^{n}_{\bfk,-\bfG_0},\label{sc_shift}
\end{align}
where the slope correction factor, $\eta$, is given by
\begin{align}
    \eta=\left(1+\frac{48e^2\theta}{\epsilon a}\sum_\bfk\frac{\Lambda^{+-}_{\bfk,-\bfG_0}\Lambda^{-+}_{\bfk,\bfG_0}}{\Delta\varepsilon_\bfk}\right)^{-1}.\label{sc_corr}
\end{align}

\noindent The interband form factors vanish at K, so the summand has no poles in the Brillouin zone, and owing to symmetries of the model is positive semi-definite. This allows us to interpret the effect of self-consistency as accounting for virtual interaction-driven interband hoppings that result from partial filling or depletion of the charge neutral system. Furthermore since $\eta<1$, this indicates a reduction of charge fluctuations and results in a more gentle filling-dependent shift in energy when self-consistency is taken into account. The fact that $\eta$ is independent of $\bfk$ indicates that self-consistency operates identically across the Brillouin zone, and suggests that $\eta$ may be thought of as an enhancement of the dielectric constant (which as we discuss elsewhere was arbitrarily imposed in the literature, even incorrectly in self-consistent calculations).

\begin{figure*}[th!]
\includegraphics[width= 0.95\linewidth]{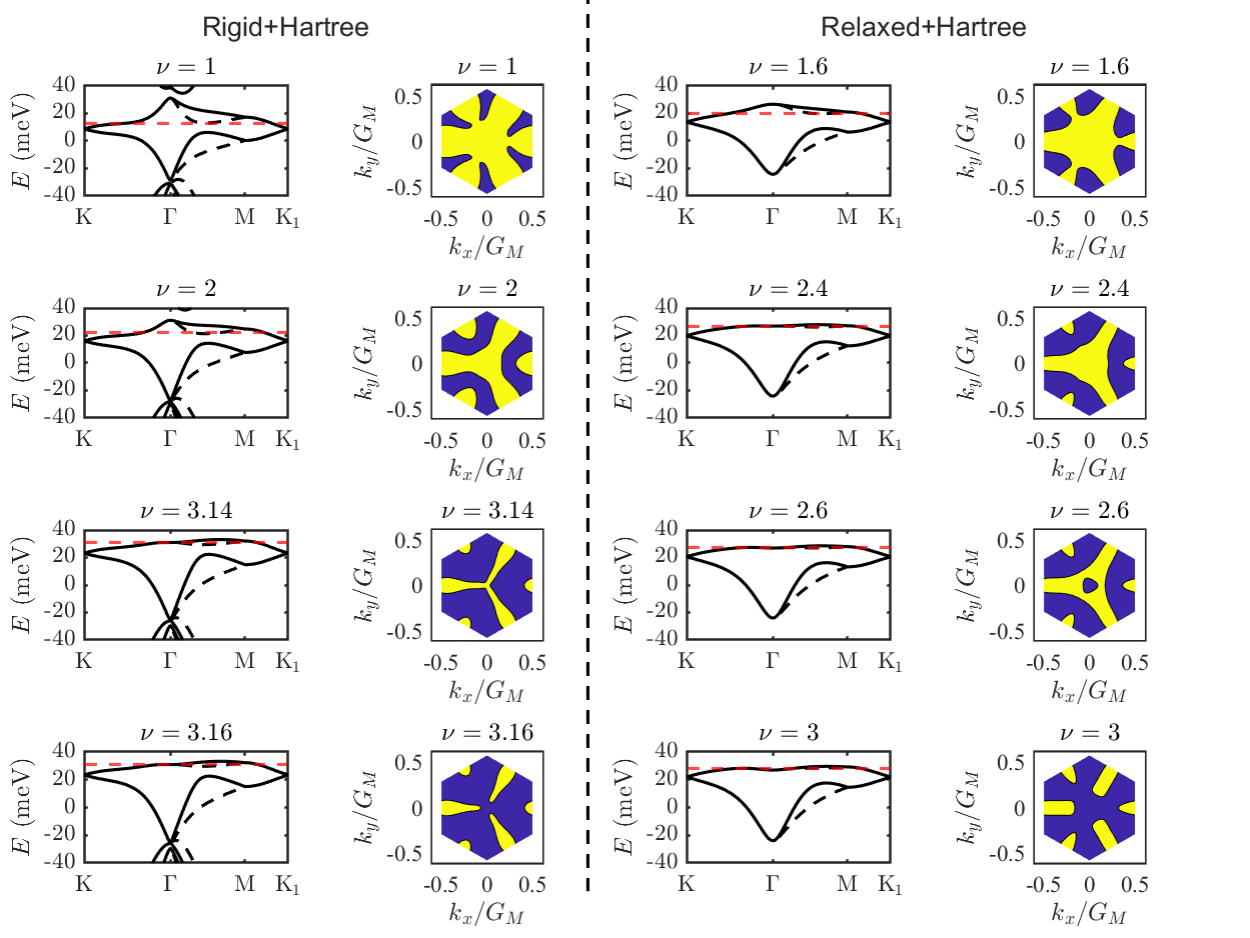}
\caption{Bandstructures and corresponding Fermi surfaces at selected fillings for the rigid (left) and relaxed (right) interacting models at $\theta=1.05$ deg. The fillings were chosen to showcase the evolution of the Fermi surface. In both models, filling starts at K and K$'$, producing a pair of Fermi surfaces. These Fermi surfaces eventually merge to produce a single Fermi surface dividing filled and unfilled regions. In the rigid model, the single Fermi surface merges with itself at the $\Gamma$ point producing three hole pockets near the three M points. In the relaxed model, an electron pocket with additional Fermi surface appears at $\Gamma$. This Lifshitz transition is not present in the rigid model, and the new Fermi surface merges with the preexisting one to form the three M pockets.}\label{Fig:fermisurface}
\end{figure*}

We see also that self-consistency becomes particularly important at angles where the noninteracting bandwidth $\Delta\varepsilon$ becomes comparable to the interaction energy scale $\frac{e^2\theta}{\epsilon a}$. This is the same result as when interactions become important in the first place, indicating that there is no regime where they may be dealt with using one-shot Hartree.  We conclude that the Coulomb interaction is either irrelevant or must be dealt with self-consistently. Finally, we note that calculation of the self-consistent correction does not itself require self-consistency, and $\eta$ is fully determined from the noninteracting model.

\section{Emergence of the Ultra Flat Band}

Our numerical results, in~\figref{os_vs_sc} indicate that at certain values of the filling factor, the energies at the M point and $\Gamma$ point cross. This in itself is unremarkable.  However, this crossing seems to coincide with a crossing of the Fermi energy, consistently no matter whether one looks at one-shot or self-consistent Hartree, and across a range of angles.  Our perturbative results, along with some numerically inspired approximations, allow us to explain this coincidence of $E_\mathrm{M}$, $E_\Gamma$ and $E_\mathrm{F}$.  First, we determine the filling $\nu^*$ at which the M and $\Gamma$ points are degenerate. The energies at these points are given by

\begin{align}
    \varepsilon_{\pm,\bfk}(\nu)&=\varepsilon_{\pm,\bfk}+\delta\varepsilon_{\pm,\bfk}(\nu)\\
    \varepsilon_{\pm,\mathrm{M}}&\approx\pm\left(\vF\frac{2\pi}{3a}\theta-W^*\right)\label{em_defn}\\
    \varepsilon_{\pm,\Gamma}&\approx\pm\left(\vF\frac{4\pi}{3a}\theta-W^*\right).\label{egamma_defn}
\end{align}
\noindent  With these expressions we may now determine the degenerate filling as

\begin{align}
    \nu^*_\pm&:\varepsilon_{\pm,\mathrm{M}}(\nu^*_\pm)=\varepsilon_{\pm,\Gamma}(\nu^*_\pm)\\
    &=\pm\frac{\pi\epsilon\vF}{9\eta e^2}\frac{1}{\Lambda^{\mathrm{sgn}(\nu^*)}_{\mathrm{K},\bfG}\left(\Lambda^\pm_{\mathrm{M},-\bfG}-\Lambda^\pm_{\Gamma,-\bfG}\right)}\label{nustar_defn}\\
    &\approx \pm\frac{\pi\epsilon\vF}{9\eta e^2\Lambda^{\mathrm{sgn}(\nu^*)}_{\mathrm{K},\bfG}\Lambda^\pm_{\mathrm{K},-\bfG}},
\end{align}
where in the last line we note that $\Lambda_\Gamma\ll\Lambda_\mathrm{M}\approx\Lambda_\mathrm{K}$, which can be seen by comparing the slopes of the energies in either~\figref{os_vs_sc} or~\figref{angle_dependence}.

We now look for the filling where the Fermi energy is degenerate with the $\Gamma$ point energy. Assuming that at this stage we are still filling carriers in the Dirac cones, the filling is given by $\nu_\Gamma=\frac{2}{\pi}A_m\kF^2$, and the degeneracy condition is written as

\begin{align}
    \pm\vF^*\kF+\delta\varepsilon_{\pm,\kF}(\nu_\Gamma)=\varepsilon_{\pm,\Gamma},
\end{align}
where $\vF^*$ is the \moire and interaction modified Dirac velocity and we explicitly take $\Lambda_\Gamma=0$. We combine this with Eqns.~\ref{sc_shift},~\ref{em_defn},~\ref{egamma_defn}, and an alternate expression for the M point energy $\varepsilon_{\pm,\mathrm{M}}\approx\pm\vF^*\frac{2\pi}{3a}\theta$ to get

\begin{align}
    \nu_\Gamma=\nu^*\left(1+\left(1-\sqrt{\frac{3\sqrt{3}\nu^*}{4\pi}}\right)\frac{\vF^*}{\vF}\right)+\mathcal{O}\left(\frac{\vF^{*2}}{\vF^2}\right),\label{nu_correction}
\end{align}
demonstrating that the three energies do indeed intersect at a single filling, up to corrections of order $\mathcal{O}(\vF^*/\vF)$.  We emphasize that these corrections are very small near the magic angle.

While the $\Gamma$ itself is largely unaffected by the Hartree interaction, the band near the $\Gamma$ point is very interesting.  It undergoes a change of curvature from negative to positive, as seen in Figs.~\ref{Fig:Bands_with_EF_rigid_Hartree} and~\ref{Fig:Bands_with_EF_relax_Hartree}, potentially resulting in an electron (hole) pocket appearing with increasing (decreasing) $\nu$, depending on whether this change of curvature occurs before or after $\Gamma$ is filled. 

Our numerical bandstructure calculations suggest that the filling at which curvature changes is close to the filling at which M and $\Gamma$ are degenerate. We may investigate the conditions for this coincidence by building on our purturbation theory results.  The effective mass at the $\Gamma$ point is determined from 

\begin{align}
    \tilde{\varepsilon}_{n,\bfk}&=\epsilon_{n,\bfk}+\eta\frac{6e^2\theta}{\epsilon a}\nu\Lambda^{\mathrm{sgn}(\nu)}_{\mathrm{K},\bfG_0}\Lambda^{n}_{\bfk,-\bfG_0}\\
    \frac{1}{\tilde{m}_n}&=\frac{1}{m^*_n}+\eta\frac{6e^2\theta}{\epsilon a}\nu\Lambda^{\mathrm{sgn}(\nu)}_{\mathrm{K},\bfG_0}\left.\left(\Lambda^{n}_{\bfk,-\bfG_0}\right)''\right|_{\bfk=\Gamma},\label{mass_renorm}
\end{align}
where $m^*_n$ and $\tilde{m}_n$ are the noninteracting and interacting effective masses at the $\Gamma$ point for band $n$ and the double prime in $\Lambda$ denotes the second momentum derivative. $m^*_+$ is negative, and $\Lambda_\bfk$ has a minimum at $\bfk=\Gamma$ and thus positive second derivative, leading to a potential sign change of $\tilde{m}_+$ for some positive $\nu=\nu_m$. We now attempt to find an expression for this filling.

As previously stated, both $\varepsilon_\bfk$ and $\Lambda_\bfk$ have local extrema at $\Gamma$, and can therefore be written

\begin{align}
    \varepsilon_\bfk&\approx\varepsilon_\Gamma+\frac{1}{2}(k-\Gamma)^2\left.\varepsilon''_\bfk\right|_{\bfk=\Gamma}\\
    \Lambda_\bfk&\approx\Lambda_\Gamma+\frac{1}{2}(k-\Gamma)^2\left.\Lambda''_\bfk\right|_{\bfk=\Gamma},
\end{align}
where we drop the band index, looking at the upper band as a specific example and ignore the anisotropy present in the system, which is unimportant for the present approximate analysis. Differentiating both equations gives

\begin{align}
    \left.\Lambda''_\bfk\right|_{\bfk=\Gamma}&=\left.\varepsilon''_\bfk\frac{\Lambda'_\bfk}{\varepsilon'_\bfk}\right|_{\bfk=\Gamma}\\
    &\approx-\frac{1}{m^*}\frac{3a}{2\pi\vF\theta}(\Lambda_\mathrm{M}-\Lambda_{\Gamma}),\label{lambdadoubleprime}
\end{align}
where in the second line we substitute the definition of $m^*$ as the second derivative of $\varepsilon$ and approximate the first derivatives of $\varepsilon$ and $\Lambda$ by a ratio of finite differences, using~\eqnref{em_defn} and~\eqnref{egamma_defn}. Substituting the result of~\eqnref{lambdadoubleprime} into~\eqnref{mass_renorm} and setting $\tilde{m}\rightarrow\infty$ then gives $\nu_m=\nu^*$, after comparison with~\eqnref{nustar_defn}. 

\begin{figure}[th!]
\includegraphics[width=1\linewidth]{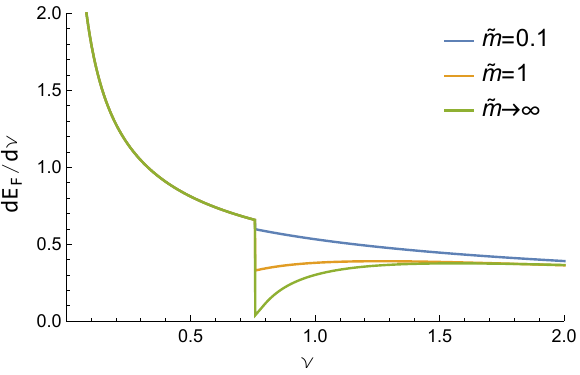}
\caption{Sensitivity of the discontinuity in $\mathrm{d}\EF/\mathrm{d}\nu$ to the interaction-modified effective mass at $\Gamma$, $\tilde{m}$. Units on both axes and of $\tilde{m}$ are arbitrary. Any positive finite $\tilde{m}$ produces a discontinuity in $\mathrm{d}\EF/\mathrm{d}\nu$, with the magnitude of the discontinuity increasing as $\tilde{m}$ increases. As $\tilde{m}\rightarrow\infty$, the minimum value of $\mathrm{d}\EF/\mathrm{d}\nu$ tends to zero.} \label{defdnu_zoom}
\end{figure}

That is, the curvature of the $\Gamma$ point vanishes at the same filling as when it is degenerate with M, for any choice of twist angle or other model parameter, provided $|\nu^*|<4$. This implies the emergence of an ultraflat band, a very large degenerate region of the Brillouin zone, and a correspondingly large density of states. 

Consequently, bearing in mind how  $\nu_m=\nu^*$ unifies two different definitions of flat bands, we are led to a single condition on whether a flat band occurs in practice; namely that the approximation 

\begin{align}\left.
    \frac{\Lambda'_\bfk}{\varepsilon'_\bfk}\right|_{\bfk=\Gamma}\approx\frac{\Lambda_\Gamma-\Lambda_\mathrm{M}}{\varepsilon_\Gamma-\varepsilon_\mathrm{M}},
\end{align}
holds. If it does, as in our relaxed interacting model, we have M being degenerate with $\Gamma$ at the same filling as when $\Gamma$ is locally flat, resulting in a wide flat region. If the approximation does not hold, as in the rigid self-consistent and either rigid or relaxed one-shot Hartree models, then M is degenerate with $\Gamma$ when $\Gamma$ is not locally flat, indicating dispersion and no wide flat region.

\section{Evolution of Band Topology as a Function of Filling}

The correction in ~\eqnref{nu_correction} depends on the values of the form factors $\Lambda$ and so depends on whether relaxation is included in the model. Specifically, we observe that the correction is negative in the rigid model and positive in the relaxed model.  This implies that the filling of the $\Gamma$ point and nearby states either precedes (rigid) or succeeds (relaxed) the band flattening; this has significant implications on the band topology as a function of filling, as seen in~\figref{Fig:fermisurface}.  We focus on the conduction band for the following discussion, as the valence band behaviour is merely reversed. In both rigid and relaxed models, filling starts at K and K$'$, generating charge pockets and a pair of Fermi surfaces. These pockets eventually merge and a single Fermi surface appears dividing occupied and unoccupied states. At this point, the topology changes; in the rigid model the $\Gamma$ point fills before it changes curvature. As such, the $\Gamma$ point remains a local maximum. As it crosses the Fermi surface, states fill from outside into $\Gamma$ and the single Fermi surface meets itself, dividing the connected hole region into three hole pockets. In the relaxed model, the $\Gamma$ point curvature flips and becomes a local minimum before it starts to fill, starting from $\Gamma$ and spreading outwards. This produces a new Fermi surface, changing the hole region from genus 0 to genus 1. The new Fermi surface then meets the old, dividing the connected hole region into three hole pockets, and the topology then continues to evolve identically in both rigid and relaxed models as the three hole pockets are depleted.

\section{Kink in $d\EF/d\nu$ and Persistence of Flat Band with Filling}

The differences in Fermi surface topology between the two models has consequences on experimental observables, particularly in the evolution of the densities of states and band energies. In the rigid model, $\nu_\Gamma<\nu^*$ and so states are filled smoothly up to $\Gamma$, resulting in a continuous $d\EF/d\nu$ seen in the second panel of Fig. 3 in the main text. In the relaxed model however, $\nu_\Gamma$ is greater than $\nu^*$ and $\nu_m$ by a small amount of order $\vF^*/\vF$ and so $\tilde{m}$ at the point where $\Gamma$ begins being filled undergoes a sign change and is enhanced over $m^*$ by an order of $-\vF/\vF^*$.

The $\Gamma$ point is thus a local minimum when it is being filled, producing a new Fermi surface and resulting in: (i) a flattening of band energies with filling, seen in Figs.~\ref{os_vs_sc}, \ref{angle_dependence} and \ref{angle_dependence2} with concomitant kinks in the densities of states, shown in the first panel of Fig. 3 in the main text as well as~\figref{flatdos}; and (ii) the discontinuity of $d\EF/d\nu$ seen in the third panel of Fig. 3 in the main text and~\figref{defdnu_zoom}. We explain these features analytically below.

\begin{figure}[th!]
\includegraphics[width=1\linewidth]{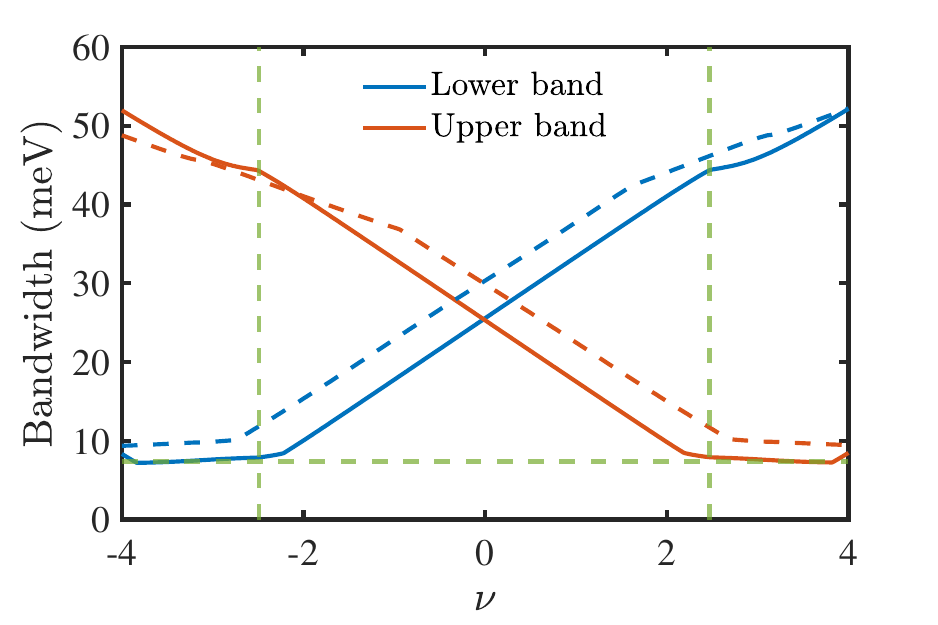}
\caption{Bandwidth of the valence (blue) and conduction (red) bands for the rigid (dotted) and relaxed (solid) interacting models. The vertical dotted green lines correspond to the kinks in energy seen in Fig. 4 of the main text, and the horizontal line marks the minimum bandwidth at that filling. The flat band exists over a wide range of fillings, with onset in the rigid (relaxed) model just after (before) $\Gamma$ gets filled.}\label{bandwidth}
\end{figure}

\begin{figure}[h!]
\includegraphics[width=1\linewidth]{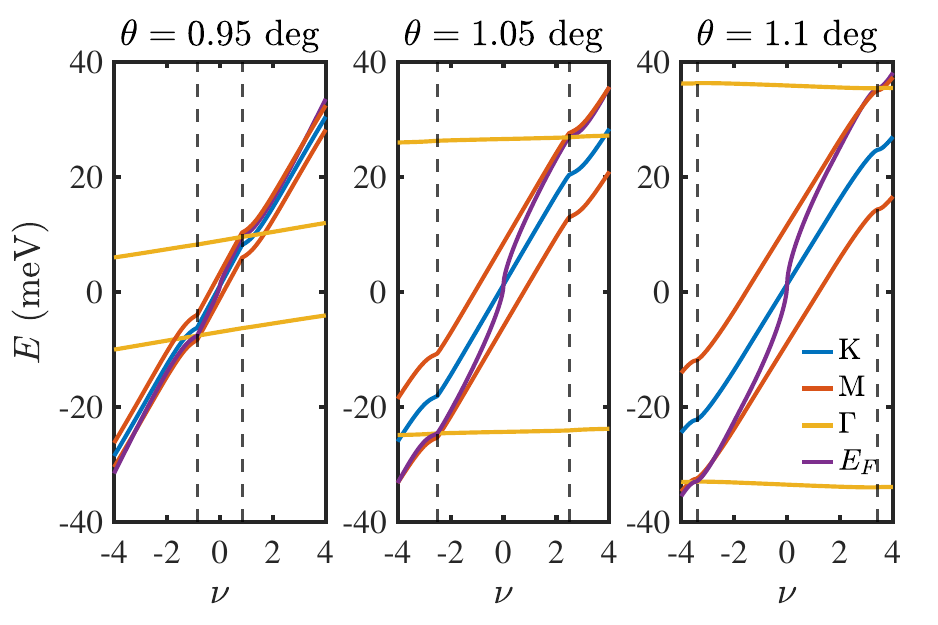}
\caption{Evolution of band energies at the K (blue), M (red) and $\Gamma$ (yellow) points, and of the Fermi energy (purple) as a function of filling in self-consistent Hartree at different twist angles. As angle increases, the linear-in-$\nu$ regime and the noninteracting bandwidth widens. The slopes of the K and M points are largely unchanged while that of $\Gamma$ changes sign, yet remaining small. The filling where the M and $\Gamma$ energies intersect increases with angle, eventually reaching unphysical values beyond $\pm4$, indicating angles for which a flat band cannot occur.}
\label{angle_dependence}
\end{figure}

\begin{figure}[th!]
\includegraphics[width=1\linewidth]{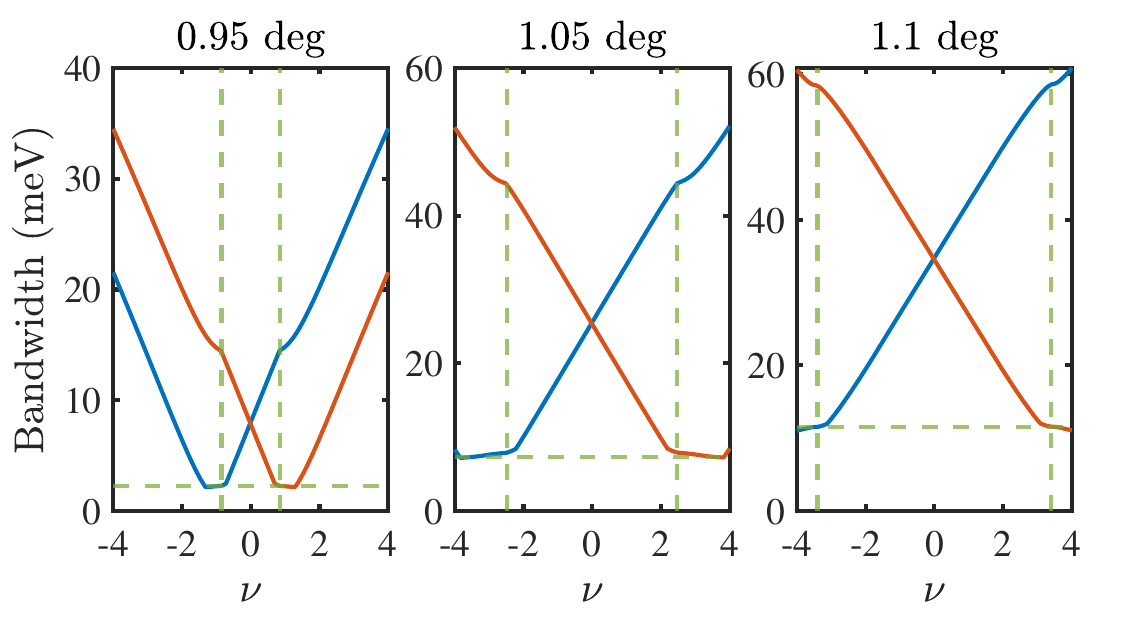}
\caption{Variation of bandwidths with angle for the valence (blue) and conduction (red) bands. The vertical dotted green lines correspond to the kinks in energy seen in~\figref{angle_dependence}, and the horizontal line marks the minimum bandwidth at that filling. Minimum bandwidth, minimum filling for minimum bandwidth and maximum filling for minimum bandwidth all increase with twist angle. As the range of physical $\nu$ is limited, the proportion of $\nu$ for which bandwidth is minimized has a maximum in twist angle.}\label{bandwidth_ang}
\end{figure}

\begin{figure}[h!]
\includegraphics[width=1\linewidth]{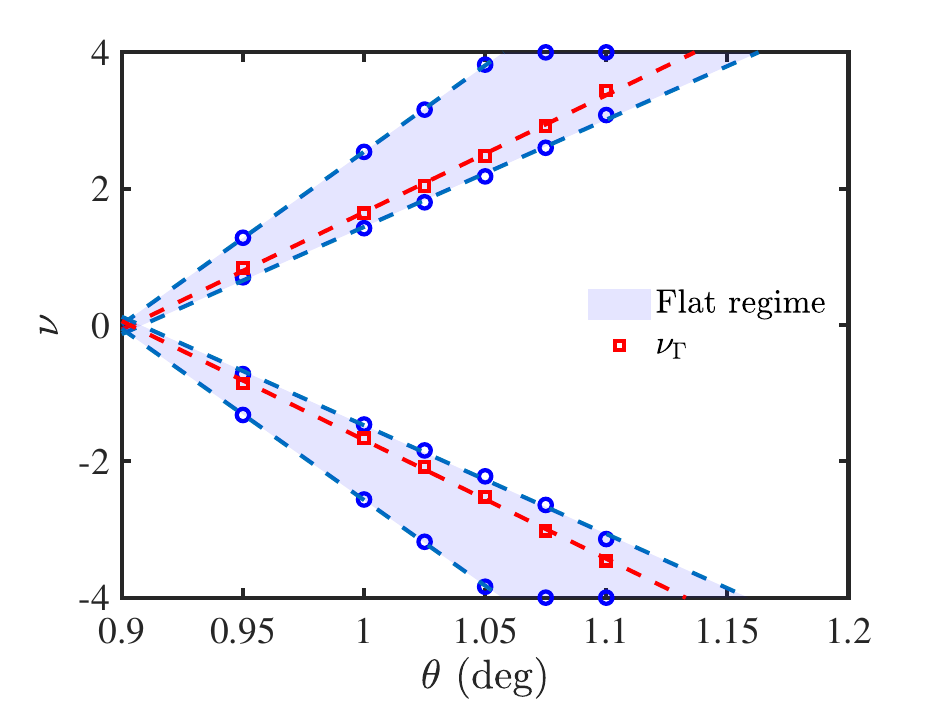}
\caption{Stability window for ultra-flat bands.  The shaded region shows the region of ultra flat band as a function of filling $\nu$ and twist angle $\theta$.  At magic angle $\theta_M \approx 0.9^\circ$, the window vanishes.  For $\theta_M < \theta \lesssim 1.05^\circ$, the fraction of band filling occupied by the ultraflat band increases reaching a maximum of about 40 percent.  The window then narrows again vanishing for $\theta > 1.15^\circ$.}
\label{angle_dependence2}
\end{figure}

We recall that one of the first approximations of the perturbative treatment was that electrons (holes) are filled (depleted) starting from the K points. For $\Gamma$ a local minimum on positive filling, a second electron pocket emerges, and the total electron density is

\begin{align}
    \rho\approx\frac{\nu_0}{4}\Lambda_\mathrm{K}+\frac{\nu_1}{4}\Lambda_\Gamma,
\end{align}
where $\nu_0$ and $\nu_1$ are the filling factors for electrons in the K and $\Gamma$ pockets respectively, and the total filling is given by $\nu=\nu_0+\nu_1$. Already we see that since $\Lambda_\Gamma\ll\Lambda_\mathrm{K}$, electrons in the K pocket contribute much more to interaction-induced band energy shifts than electrons in the $\Gamma$ pocket.

We now split up the Brillouin zone, denoting momenta near the K points with $\bfk$ as we have done till now, and momenta near the $\Gamma$ points with $\bfq$, to get the approximate interaction-modified bands as

\begin{align}
    \tilde{\varepsilon}_\bfk&\approx\vF^*k+\xi_\mathrm{K}\nu_0\\
    \tilde{\varepsilon}_\bfq&\approx\varepsilon_\Gamma+\left(\frac{\nu_0}{\nu^*}-1\right)\frac{q^2}{2|m^*|}+\frac{q^4}{\sigma}+\xi_\mathrm{\Gamma}\nu_0,
\end{align}
where we use the shorthand $\xi_\bfk=6\eta U_{\bfG_0}\Lambda_\mathrm{K}\Lambda_\bfk$ to simplify the expression for the interaction correction, and we remind the reader that $m^*$ is negative. We also take the aforementioned simplification that interaction corrections to either the band energies or the effective mass are due to $\nu_0$ only, and expand to 4\textsuperscript{th} order in $q$ as we are interested in $\nu$ just above when $\Gamma$ starts being filled, where the interaction-modified effective mass diverges. The quantity $\sigma$ has units of areal mass density. Noting that both pockets are filled up to a common Fermi energy, $\EF=\tilde{\varepsilon}_{\kF}=\tilde{\varepsilon}_{q_\mathrm{F}}$, we use the expressions for the filling in both pockets, $\nu_0=\frac{2}{\pi}A_m\kF^2$ and $\nu_1=\frac{1}{\pi}A_m q_\mathrm{F}^2$ and the definition of total filling to determine how the interaction-relevant filling $\nu_0$ varies with total filling $\nu$ for $\nu>\nu_\Gamma$ i.e. when the $\Gamma$ point starts being filled. The factor of 2 difference in the definitions of $\nu_0$ and $\nu_1$ in terms of $\kF$ and $q_\mathrm{F}$ is due to there being two K pockets about K and K$'$, and only a single $\Gamma$ pocket. The expression for the interaction-relevant filling is then

\begin{align}
    \nu_0-\nu_\Gamma=\frac{\pi\nu_\Gamma}{A_m\vF k_\mathrm{M}}\left(\frac{1}{2\tilde{m}}(\nu-\nu_\Gamma)+\frac{\pi}{\sigma A_m}(\nu-\nu_\Gamma)^2\right),\label{active_nu}
\end{align}
where $k_\mathrm{M}=2\pi\theta/3a$ is the momentum difference between K and M points. When $\Gamma$ is just being filled and $1/\tilde{m}$ is very small, of order $\vF^*/\vF$, the portion of filling relevant to interactions $\nu_0$ varies quadratically with total filling $\nu$, leading to a a sudden kink in the band energies with $\nu$, as seen in Figs~\ref{os_vs_sc} and~\ref{angle_dependence}. This universal rescaling of ``active'' filling fraction from $\nu$ to $\nu_0$ results in a kink in the evolution of all band energies as a function of total filling $\nu$, leading to kinks in the van Hove singularity energies as seen in~\figref{flatdos}.

With~\eqnref{active_nu}, we may express the Fermi energy as a function of total filling $\nu$ and thus obtain $\mathrm{d}\EF/\mathrm{d}\nu$, with values determined from the real bands shown in the second panel of main Fig. 3 and qualitative behaviour on changing $\tilde{m}$ shown in~\figref{defdnu_zoom}. The initial $1/\sqrt{\nu}$ behaviour of $\mathrm{d}\EF/\mathrm{d}\nu$ is due to the Dirac cone at K, and band energies changing to first order in $k$. When the $\Gamma$ point is filled, the new electron pocket causes a discontinuity in $\mathrm{d}\EF/\mathrm{d}\nu$ whose magnitude depends on $\tilde{m}$, that is on the magnitude of the rescaled linear dependence of $\nu_0$ on $\nu$ away from unity. For diverging $\tilde{m}$, $\mathrm{d}\EF/\mathrm{d}\nu$ falls to zero as the entire band, and thus $\EF$, evolves as a quadratic minimum with $\nu$. The qualitative behaviour of this diverging mass limit matches well with the numerical results of the full model.

Another consequence of the kink in $\mathrm{d}\EF/\mathrm{d}\nu$ is the persistence of minimum bandwidth with $\nu$.  Since we have determined~\eqnref{active_nu} defining the active filling fraction $\nu_0$, and thus the interaction corrected bands for all total filling fractions $\nu$, we are now equipped to explain the evolution of the bandwidth, shown in Figs.~\ref{bandwidth} and~\ref{bandwidth_ang}. We see that a nonzero minimum in bandwidth exists and, more importantly, is maintained over a range of $\nu$. Furthermore, the onset of the minimum bandwidth is slightly before $\nu_\Gamma$ in the relaxed model and after it in the rigid model. The onset is in fact set by $\nu^*$, with the deviations of either model seen in~\figref{bandwidth} corresponding to the sign of the order $\vF^*/\vF$ correction of $\nu_\Gamma$ compared to $\nu^*$. Inspection of the bandstructures in~\figref{Fig:Bands_with_EF_relax_Hartree} indicate that bandwidth is well approximated by the largest pairwise difference between $\tilde{\varepsilon}_\mathrm{K}$, $\tilde{\varepsilon}_\mathrm{M}$ and $\tilde{\varepsilon}_\Gamma$. This matches our previous argument, with the onset of minimum bandwidth occuring when M is degenerate with $\Gamma$, that is at $\nu=\nu^*$. 

For $\nu>\nu^*$, the K and M energies continue to increase while the $\Gamma$ energy hardly moves, remaining between the two. The bandwidth is then characterized by $\tilde{\varepsilon}_\mathrm{M}-\tilde{\varepsilon}_\mathrm{K}\approx\varepsilon_\mathrm{M}-\varepsilon_\mathrm{K}$, where we recall that $\Lambda_\mathrm{K}\approx\Lambda_\mathrm{M}$ and so the band energies evolve identically with filling, even after $\Gamma$ starts being filled. This minimum bandwidth is thus maintained up to a maximum filling $\nu_\mathrm{max}$ given by either by the intersection of the K energy with the $\Gamma$ energy, or by the maximum physical value of 4,

\begin{align}
    \nu_\mathrm{max}=\mathrm{min}\left(\nu_\Gamma+\frac{A_m}{\pi}\sqrt{\sigma\varepsilon_\mathrm{M}},4\right),
\end{align}
where the noninteracting M point energy appears because $\epsilon_\Gamma-\vF k_\mathrm{M}=\epsilon_\mathrm{M}$, and $\vF k_\mathrm{M}$ appears in~\eqnref{active_nu} relating $\nu_0$ to $\nu$. If the region of minimum bandwidth is not cut off by the hard physical limit of $\nu=4$, the length of the region is given simply by

\begin{align}
    \Delta\nu=\frac{A_m}{\pi}\sqrt{\sigma\varepsilon_\mathrm{M}}+\mathcal{O}\left(\frac{\vF^*}{\vF}\right).\label{flatwidth}
\end{align}

\section{Stability of heavy fermion pocket and ultraflat band}

Equation~\ref{nu_correction} seems to imply that any model of interacting twisted bilayer graphene should have $\nu_\Gamma\approx\nu^*$ and thus very flat bands on the verge of being filled near the magic angle where $\vF^*/\vF$ is minimized, yet we know this is not true in, for example, the one-shot Hartree result, or in models without full lattice relaxation \cite{guinea_electrostatic_2018, rademaker_charge_2019, cea2019electronic,xie2020nature, goodwin_hartree_2020, zhang_correlated_2020, bultinck_ground_2020, lewandowski_does_2021, liu_theories_2021, bernevig_twisted_2021, song2022magic, wagner_global_2022}.  It is therefore necessary to examine further the coefficient of the $\mathcal{O}(\vF^*/\vF)$ term more closely.  We can write this term as $(1-\sqrt{\nu^*/\nu_\mathrm{c}})$, with some critical filling $\nu_\mathrm{c}$. The value of $\nu^*$ in comparison with this critical value will then determine how close, proportionally, $\nu_\Gamma$ gets to $\nu^*$, and whether $\nu_\Gamma < \nu^*$ as in the rigid approximation or $\nu_\Gamma > \nu^*$ as found in the relaxed case.  The simplifying approximations we have used above for the analytical model give $\nu_\mathrm{c}=4\pi/3\sqrt{3}\approx2.42$ while at $\theta=1.05^\circ$ we have $\nu^* = 2.36<\nu_\mathrm{c}$, giving a very small, positive correction. In other models such as Ref.~\cite{zhu2024gw}, where $\nu^*$ is relatively small, the relative distance between $\nu_\Gamma$ and $\nu^*$ widens, and consequently ultra flat bands are not relevant, appearing at values of $\nu$ far from where the $\Gamma$ point is filled.

We now attempt to determine the twist angle dependence of the quantities derived above, in particular $\nu^*$ and $\nu_\mathrm{max}$ highlighted in Figs.~\ref{angle_dependence} and~\ref{bandwidth_ang} respectively. The explicit angle dependence of $U_\bfG$ is already shown in~\eqnref{sc_shift} so what remains is to determine the $\theta$ dependence of $\nu$, $\Lambda$ and $\eta$. We note here that the following arguments will be most relevant above magic angle, where the band behaviour is monotonic.

We first note that $\nu$ is defined as the number of electrons per \moire unit cell, and as such is independent of $\theta$. On the other hand, $\rho$ is the number of electons per unit area, and therefore changing $\theta$, and thus the area of the \moire cell, at fixed $\nu$ results in changes to $\rho$. From~\eqnref{os_rhog}, this means $\Lambda\sim\mathcal{O}(\frac{1}{\theta^2})$, as expected since $\Lambda$ measures the overlap between states related by a \moire reciprocal lattice vector, which is expected to decay as $\theta$ increases and the \moire reciprocal lattice vectors get longer. The separation between conduction and valence band $\Delta\varepsilon$ increases linearly with $\theta$, and so the self-consistent correction $\eta-1\sim\mathcal{O}(\frac{1}{\theta^4})$ for large $\theta$, indicating a vanishing importance of self-consistency at large twists, again as expected. The energy shift therefore varies as $\delta\varepsilon_{n,\bfk}(\theta,\nu)\sim\mathcal{O}(\frac{1}{\theta^3})$, and the effect of interactions as a whole vanishes at high twists. 

\begin{figure}[ht!]
\includegraphics[width=0.9\linewidth]{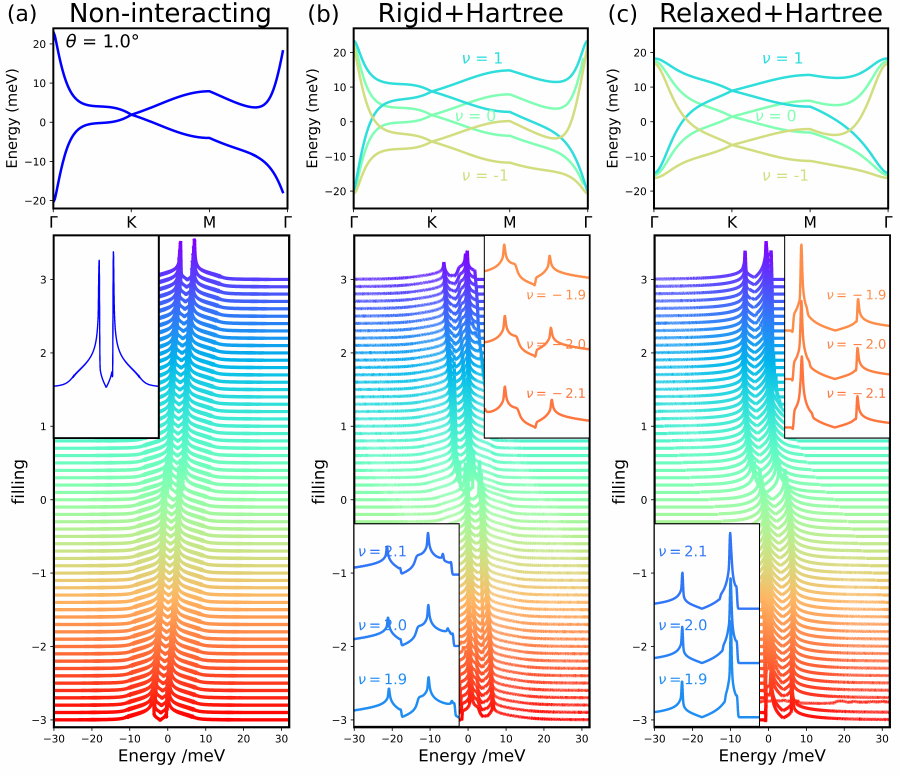}
\caption{Simulation of scanning probe measurements.  Top panel shows band structures and lower panel shows density of states for the noninteracting (left), rigid interacting (middle) and relaxed interacting (right) models. Interactions introduce a filling dependence to the band structures.  In addition to restoring the experimentally observed two-peak structure, relaxation also widens the vHS separation to $\approx7$ meV, consistent with experiments. The Lifshitz transition is not visible in such measurements.}
\label{Fig:DOS}
\end{figure}

\begin{figure}[h!]
\includegraphics[width=1\linewidth]{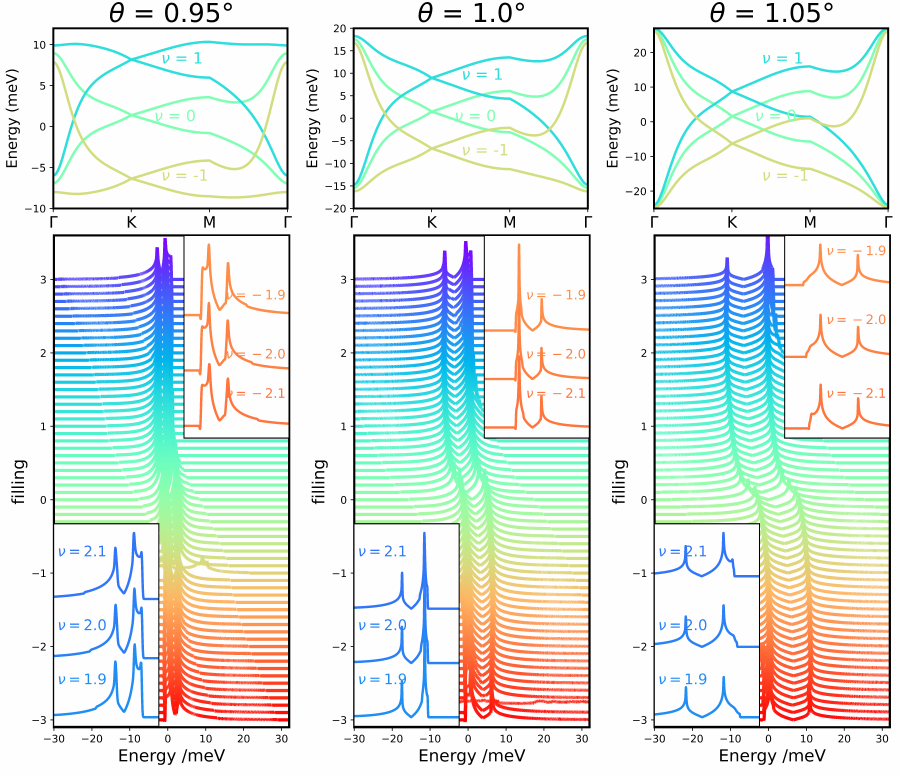}
\caption{Bandstructures (top) and density of states (bottom) for the relaxed interacting model at various twist angles. Beyond the van Hove singularity separation increasing with twist angle as anticipated in the non-interacting model, the kink feature at $\nu^*$ in tunneling density of states moves to higher filling as discussed in the main text.  Similar to Fig.~\protect{\ref{angle_dependence}}, for $\theta = 0.95^\circ$, one can observe the pinning and unpinning of the flat band.\label{flatdos}
}
\end{figure}

From~\eqnref{nustar_defn} we see immediately that $\nu^*\sim\mathcal{O}(\theta^3)$, corroborating \figref{angle_dependence} which shows that the fillings at which the M and $\Gamma$ points are degenerate increase with twist angle. $\nu_\Gamma$, being approximately equal to $\nu^*$, behaves similarly. The minimum bandwidth is given by $\varepsilon_\mathrm{M}-\varepsilon_\mathrm{K}\sim\mathcal{O}(\theta)$, an increase as seen in~\figref{bandwidth_ang}. Finally, $\sigma$ evolves as $q^4$ and so the leading order term term of~\eqnref{flatwidth} does not vary with angle. The coefficient of the correction term likewise does not vary and so the angle dependence is controlled by $\vF^*/\vF\sim\mathcal{O}(\theta)$, an increase with $\theta$, seen in the left and middle panels of~\figref{bandwidth_ang}, before $\nu_\mathrm{max}$ is cut off.

\section{STM Experiments \label{Sec:STM}}

In this section we simulate scanning probe spectroscopy.  Since the kink equally shifts the Fermi energy and the $M$-point, the separation of the van Hove singularities will be unchanged by the kink and emergence of the heavy fermion.  This can be seen in Fig.~\ref{Fig:DOS} where the Self-consistent Hartree model for the rigid model and that of the relaxed model look qualitatively similar despite the heavy fermion and flat band emerging in the latter.  Similar conclusions can be gleaned from the angle dependence seen in Fig.~\ref{flatdos}.  Despite not being able to observe the kink, we can still make a few useful observations.  Unlike the rigid model where Hartree calculations give spurious additional peaks (middle panel of Fig.~\ref{Fig:DOS} that is consistent with Ref.~\cite{cea2019electronic}), the density of states for the relaxed model has a clean two peak structure with strong asymmetry in the magnitude of the two peaks. These are experimentally observable features that emerge naturally in our self-consistent Hartree model with relaxation and with no fine tuning of any parameters. \\

\begin{figure}[ht!]
\centering
\begin{subfigure}[b]{0.46\textwidth}
    \centering
    \includegraphics[width= \linewidth]{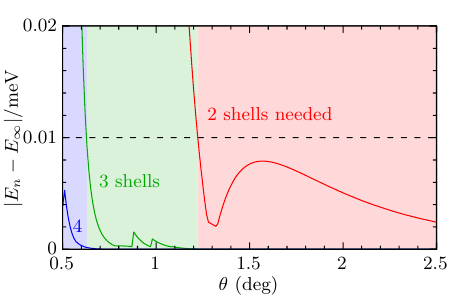}
\end{subfigure}
\begin{subfigure}[b]{0.5\textwidth}
    \centering
    \includegraphics[width= \linewidth]{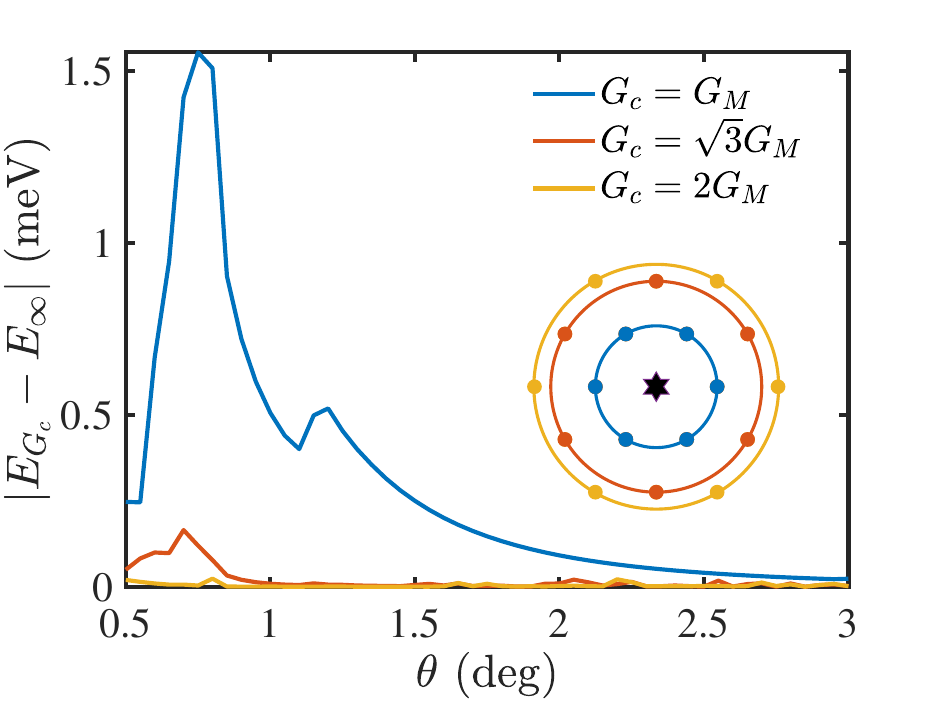}
\end{subfigure}
\caption{Benchmarking results of the number of cutoff shells for the continuum model (top) and for self-consistent Hartree bands (bottom). The error here is defined as the maximum energy difference between two low-energy bands compared with reference bands. The reference bands have been chosen with a sufficiently large number of cutoff shells. The inset of the bottom panel presents the reciprocal lattice vector included for a given cutoff. To achieve enough precision, a large number of cutoffs is required. In this work, we use 4 shells for the continuum model and set $G_c=2G_M$ for the Hartree calculation to ensure convergence across all ranges of twist angle.}
\label{fig:Benchmarking1}
\end{figure}

\section{\label{Sec:Benchmarking} Numerical Benchmarking}
This section provides benchmarking results for all parameters and approaches used in the main text. The goals of this section are to (i) check and verify the numerical convergence for the parameters we use, and (ii) provide additional information to future works regarding the minimal cutoff required to achieve a certain accuracy.

\begin{figure*}[th!]
\includegraphics[width= \linewidth]{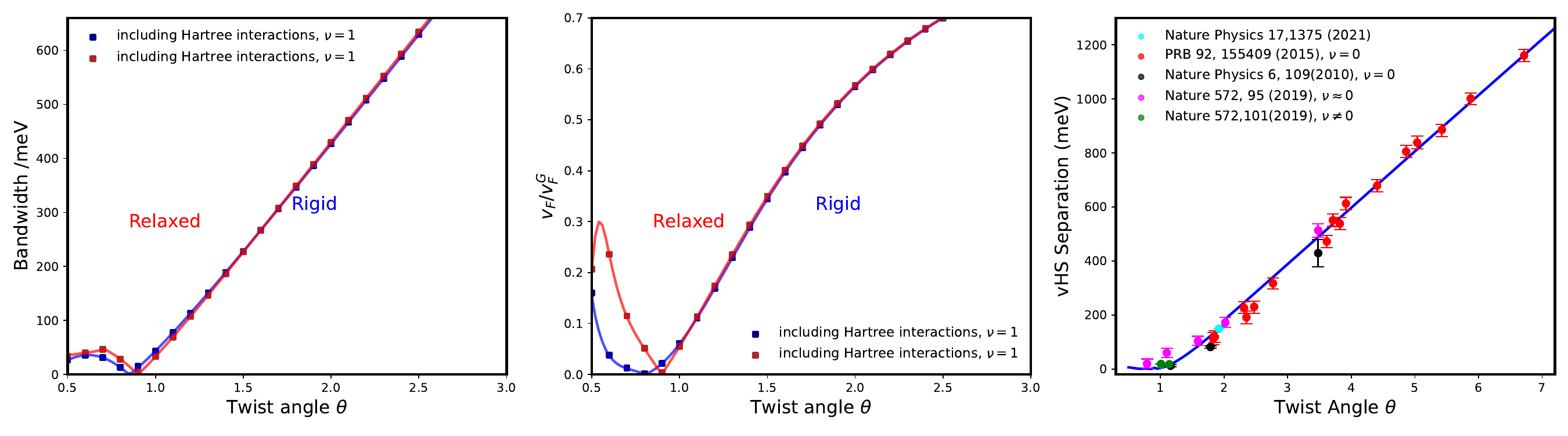}
\caption{ Validation of the large-angle continuum model. Left and middle panels: Examination of bandwidth and Fermi velocity across various twist angles. The red (blue) line represents the continuum model with (without) relaxation, incorporating Hartree self-energy denoted by corresponding dots. In the large-angle regime, the effects of both relaxation and Hartree effects become negligible. Right panel: Comparison of van Hove singularities separation in the relaxed model with experimental STM data \cite{choi2021interaction,wong2015local,li2010observation,kerelsky2019maximized,xie2019spectroscopic}. The continuum model's prediction aligns well with experimental data in the large-angle region. However, the uncertainty in experimental measurements is considerable near the magic angle, making it challenging to draw solid conclusions.}
\label{FIG:Validation}
\end{figure*}

\subsection{Numerical Convergence for Number of Continuum Model Cutoff Shells and Number of Low Energy Hartree Bands}

The benchmark results of the cutoff of the continuum model and Hartree self-energy are summarized in Fig.~(\ref{fig:Benchmarking1}). Here, we define the error as the maximum energy difference between two low-energy bands compared with reference bands. The reference bands have been chosen with a sufficiently large number of shells. To achieve an error smaller than 0.01~$\mathrm{meV}$, at least 2 shells are needed for twist angles $\theta\gtrsim1.2~\mathrm{deg}$, and 3 shells are needed for $\theta\gtrsim0.6~\mathrm{deg}$. Similarly, the numerical error for Hartree cutoff shells is shown in the bottom panel. The inset in this panel shows the reciprocal lattice vector we in-
cluded for a given cutoff. To ensure convergence for all twist angles investigated, we choose the cutoff by including the 4th shell for the continuum model and set $G_c=2G_M$ for the Hartree calculation.  For the Hartree calculations, we use 248 valence and conduction bands with 40*40=1600 k-points in the MBZ. In some calculations which require higher accuracy, like Fig.4 in the main text, we use 60*60=3600 k-points in the MBZ.

\begin{figure}[h!]
\includegraphics[trim=3.75cm 3cm 3cm 3cm, width= 1\linewidth]{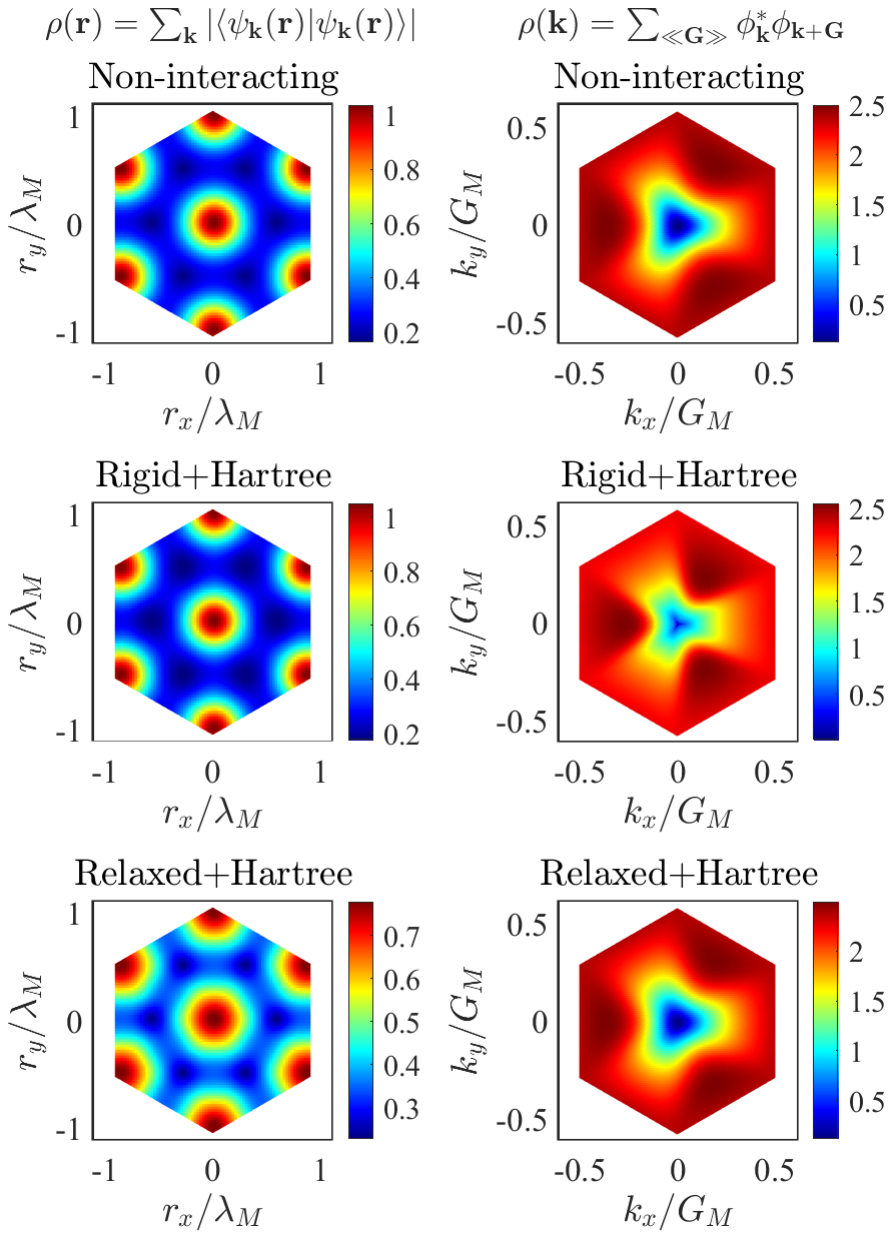}
\caption{Charge density in real space (left column) and momentum space (right column) at $\theta=1.05\ \mathrm{deg}$. In both rigid/relaxed with Hartree interaction cases, we utilize the filling factor $\nu=1$. In real space, charge accumulates in well-defined, localized regions corresponding to minima of the \moire potential, forming a triangular lattice. In momentum space, the sum over reciprocal lattice density harmonics is peaked at the K points and vanishes at the $\Gamma$-point, resulting in corresponding shifts of the bands as seen in Fig.~(\ref{Fig:ShiftOfBands}) of the main text.}
\label{fig:FormFactorChargeDistribution}
\end{figure}

\begin{figure}[h!]
\includegraphics[width=1\linewidth]{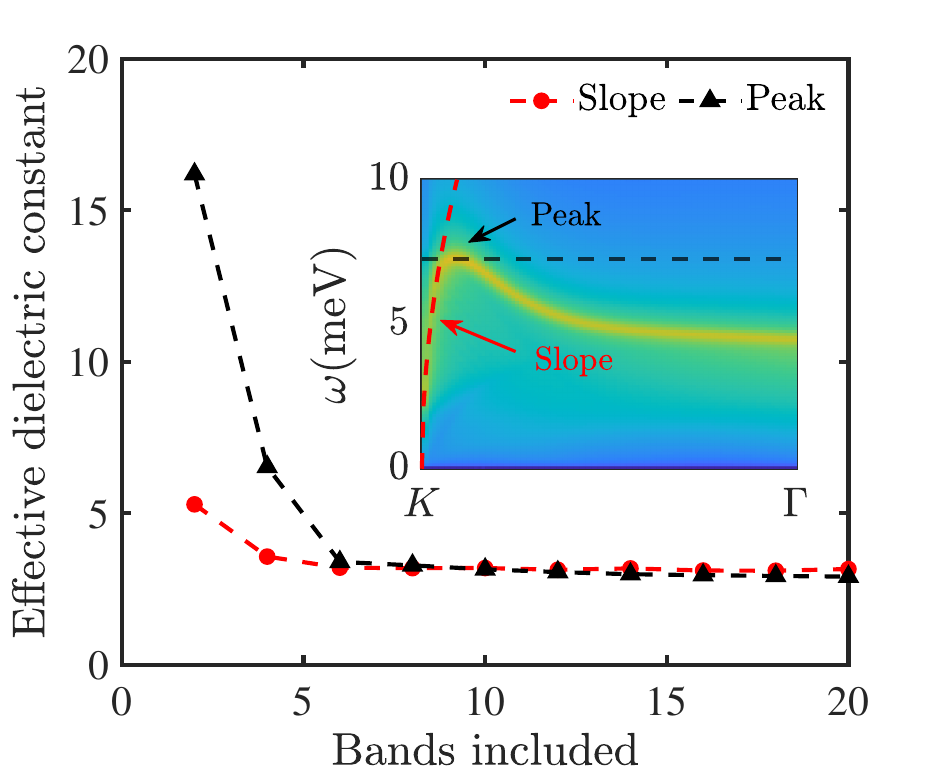}
\caption{Effective dielectric constant as a function of bands included by comparing both slope and peak of plasmon dispersion. For a fixed number of low-energy bands, the plasmon dispersion is solely determined by the dielectric constant. The effective dielectric constant can be found by matching the slope or peak of the plasmon dispersion, as shown in the inset, with the reference plasmon dispersion. Here, the reference plasmon dispersion is defined when we include all energy bands. We find that the effective dielectric constant becomes closer to the true dielectric constant as we include more bands, which implies that once higher bands are included in the calculation there is no need to artificially enhance the dielectric constant.}
\label{fig:EffDielectricConst}
\end{figure}

\begin{figure}[h!]
\centering
\begin{subfigure}[b]{0.48\textwidth}
    \centering
    \includegraphics[width= \linewidth]{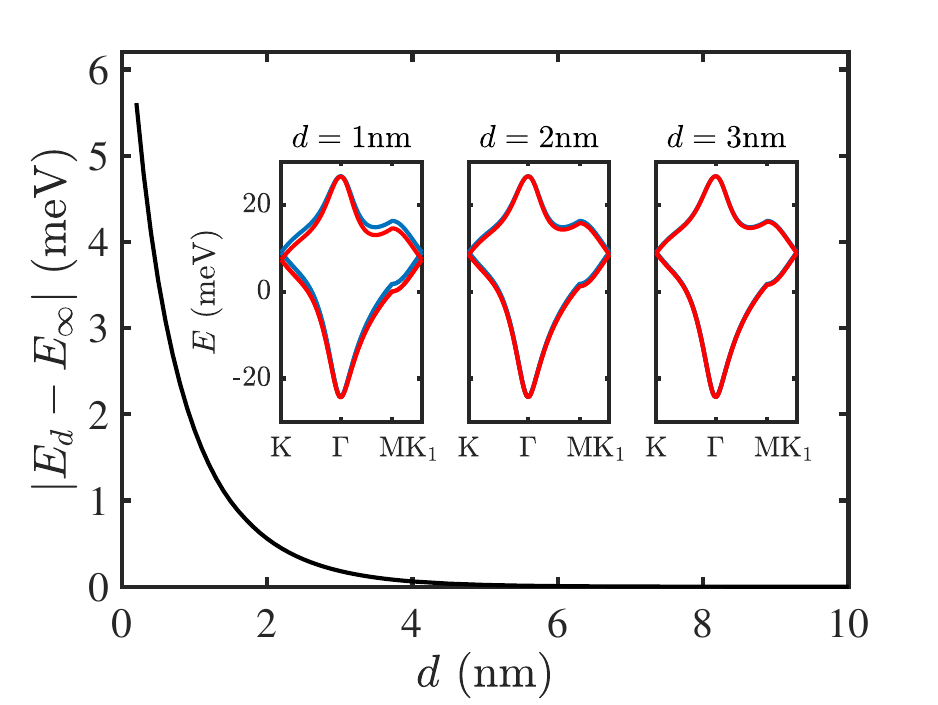}
\end{subfigure}
\begin{subfigure}[b]{0.48\textwidth}
    \centering
    \includegraphics[width= \linewidth]{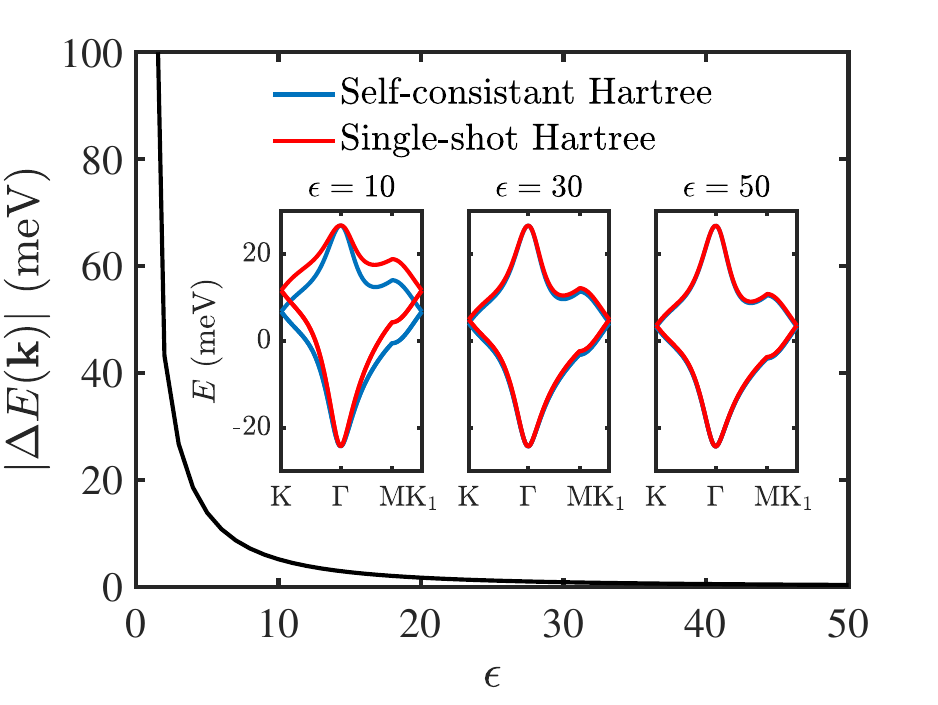}
\end{subfigure}
\caption{Maximum energy difference between two low-energy bands for (top): different metal gate separation distances, and (bottom): the one-shot versus self-consistent calculation. In the top panel, $d\rightarrow\infty$ has been chosen as the reference point. The inset in this panel compares the bands when $d\rightarrow\infty$ (blue) with $d=1,2,3~\mathrm{nm}$ (red). One can find that the metal gate becomes irrelevant when $d\gtrsim2~\mathrm{nm}$. Similarly, the bottom panel benchmarks the different dielectric constants $\epsilon$ for both single-slot and self-consistent Hartree band structures, with the inset showing band structures for different dielectric constants. Both approaches start to deviate when $\epsilon\lesssim15$.}
\label{fig:Benchmarking2}
\end{figure}

\subsection{Validation of the Large Angle Continuum Model}

We begin by validating the large-angle continuum model, with the benchmarking results summarized in Fig.~\ref{FIG:Validation}. Here, we have computed three key quantities predicted by the continuum model: (i) bandwidth, (ii) Fermi velocity, and (iii) van Hove singularities separation. The red (blue) line represents the continuum model with (without) relaxation, incorporating Hartree self-energy denoted by corresponding dots. As mentioned in the main text, both relaxation and Hartree interaction become significant only near the magic angle. Therefore, the bandwidth and Fermi velocity for all four cases collapses with each other at large twist angles. In the small angle regime, relaxation acts as a shift of the "magic angle." Contrary to expectations, we found that Hartree interaction does not alter the bandwidth and Fermi velocity, explaining why Hartree is not detected in typical transport or STM experiments. 

In the right panel of Fig.~\ref{FIG:Validation}, we compare the predicted van Hove singularities separation across a wide range of twist angles with experimental STM data. The predictions from the continuum model agree well with experimental data in the large-angle region. However, the uncertainty in experimental measurements is considerable near the magic angle, making it challenging to draw solid conclusions.

\subsection{Evolution of the Real Space Density and Momentum Space Form Factors with Relaxation and Hartree Interactions}

Figure~(\ref{fig:FormFactorChargeDistribution}) illustrates the charge density in real space and momentum space at $\theta=1.05\ \mathrm{deg}$. Here we compare the non-interacting case with both rigid and relaxed states incorporating Hartree interaction. In real space, charge accumulates in well-defined, localized regions corresponding to minima of the \moire potential, resulting in a triangular lattice Wigner crystal of \moire atoms. The corresponding momentum distribution exhibits high anisotropy, leading to shifts of the bands as observed in Fig.~(\ref{Fig:ShiftOfBands}) in the main text. Numerical results further demonstrate that the charge distribution becomes narrower in the rigid model while wider in the relaxed model.

We note that the charge density here is associated with the expansion coefficients of the Hartree potential defined in Eq.~(\ref{eq:coeff_HTpotential}), which are obtained by summing over all sublattice indices and reciprocal lattice vectors. It's important to recognize that the regime below the Fermi surface for charge density directly contributes to the Hartree potential. This verifies that the contribution of charge density to the Hartree potential is non-uniform.

\subsection{Effective Random Phase Approximation Dielectric Constant as a Function of Included Bands}

The idea of effective random phase approximation dielectric constant was first proposed by Ref~\cite{lewandowski2019intrinsically}, where they found a higher dielectric constant is needed to match the plasmon dispersion of a two-bands top model with a full band continuum model. As shown in Fig.~\ref{fig:EffDielectricConst}, insofar as the plasmon dispersion is concerned, including only two bands requires an enhancement of $\epsilon$ by a factor ranging between $\sim 5$ to $\sim 15$.  Based on this observation, several works in the literature proceeded to do Hartree calculations based on two low-energy bands with a large dielectric constant.  However, as can also be seen in Fig.~\ref{fig:EffDielectricConst}, the effective dielectric constant becomes closer to the true dielectric constant as we include more bands.  We conclude that the artificially large dielectric is unphysical, and is only a result of inappropriately truncating the low-energy bands.  The physically relevant dielectric constant with the full band model is the correct way to include Hartree self-energy. \\

\subsection{Numerical Convergence for Metal Gate Separation $d$ and Dielectric Constant $\epsilon$}

Finally, Fig.(\ref{fig:Benchmarking2}) shows the role of external metal gate and the differences between single-shot and self-consistent Hartree calculation. As we mentioned in Sec.\ref{sm_mp}, Coulomb interaction $V(q,\epsilon,d)$ is in general included the distance $d$ from the metallic screening gate. However, we find that the metallic screening gate becomes irrelevant when $d\gtrsim2~\mathrm{nm}$, which is much smaller than the typical distance in experiment. Therefore, we adopt the limit $d\rightarrow\infty$ in the main text for simplicity. We also find that the single-shot and self-consistent Hartree provide different band structures when $\epsilon\lesssim15$. To obtain accurate bands, one needs to do the self-consistent calculation due to the small dielectric constant in the experimental setup (typically $\epsilon=4$ for SiO$_2$ and hBN substrates).

\end{document}